\documentclass[apj]{emulateapj}
\newcommand\submitms{n}		
\if\submitms y
\else
\usepackage{apjfonts}
\fi

\usepackage{ifthen}
\usepackage{natbib}

\if\submitms y
\bibpunct[, ]{(}{)}{,}{a}{}{,}
\else
\fi

\shorttitle{Spitzer Secondary Eclipses of WASP-18b}
\shortauthors{Nymeyer {\em et al.}}

\newcommand\degree{\degr}
\newcommand\degrees\degree

\newcounter{fignum}

\DeclareSymbolFont{UPM}{U}{eur}{m}{n}
\DeclareMathSymbol{\umu}{0}{UPM}{"16}
\let\oldumu=\umu
\renewcommand\umu{\ifmmode\oldumu\else\math{\oldumu}\fi}
\newcommand\micro{\umu}
\renewcommand\micron{\micro m}
\newcommand\microns \micron

\let\oldsim=\sim
\renewcommand\sim{\ifmmode\oldsim\else\math{\oldsim}\fi}
\let\oldpm=\pm
\renewcommand\pm{\ifmmode\oldpm\else\math{\oldpm}\fi}
\newcommand\by{\ifmmode\times\else\math{\times}\fi}
\newcommand\ttt[1]{10\sp{#1}}

\newcommand\tablebox[1]{\begin{tabular}[t]{@{}l@{}}#1\end{tabular}}
\newbox{\wdbox}
\renewcommand\c{\setbox\wdbox=\hbox{,}\hspace{\wd\wdbox}}
\renewcommand\i{\setbox\wdbox=\hbox{i}\hspace{\wd\wdbox}}

\newcommand\herenote[1]{{\bfseries #1}\typeout{======================> note on page \arabic{page} <====================}}

\newcount\timect
\newcount\hourct
\newcount\minct
\newcommand\now{\timect=\time \divide\timect by 60
         \hourct=\timect \multiply\hourct by 60
         \minct=\time \advance\minct by -\hourct
         \number\timect:\ifnum \minct < 10 0\fi\number\minct}

\newcommand\mctc{\multicolumn{2}{c}}

\catcode`@=11

\if\submitms y
\renewcommand\comment[1]{}
\else
\newcommand\comment[1]{}
\fi
\newcommand\commenton{\catcode`\%=14}
\newcommand\commentoff{\catcode`\%=12}

\renewcommand\math[1]{$#1$}
\newcommand\mathshifton{\catcode`\$=3}
\newcommand\mathshiftoff{\catcode`\$=12}

\comment{the backslash is necessary}

\comment{alignment tab}

\let\atab=&
\newcommand\atabon{\catcode`\&=4}
\newcommand\ataboff{\catcode`\&=12}

\let\oldmsp=\sp
\let\oldmsb=\sb
\def\sp#1{\ifmmode
           \oldmsp{#1}%
         \else\strut\raise.85ex\hbox{\scriptsize #1}\fi}
\def\sb#1{\ifmmode
           \oldmsb{#1}%
         \else\strut\raise-.54ex\hbox{\scriptsize #1}\fi}
\newbox\@sp
\newbox\@sb
\def\sbp#1#2{\ifmmode%
           \oldmsb{#1}\oldmsp{#2}%
         \else
           \setbox\@sb=\hbox{\sb{#1}}%
           \setbox\@sp=\hbox{\sp{#2}}%
           \rlap{\copy\@sb}\copy\@sp
           \ifdim \wd\@sb >\wd\@sp
             \hskip -\wd\@sp \hskip \wd\@sb
           \fi
        \fi}
\def\msp#1{\ifmmode
           \oldmsp{#1}
         \else \math{\oldmsp{#1}}\fi}
\def\msb#1{\ifmmode
           \oldmsb{#1}
         \else \math{\oldmsb{#1}}\fi}
\def\supon{\catcode`\^=7}
\def\supoff{\catcode`\^=12}
\def\subon{\catcode`\_=8}
\def\suboff{\catcode`\_=12}
\def\supsubon{\supon \subon}
\def\supsuboff{\supoff \suboff}

\comment{
\let\oldmsp=\sp
\let\oldmsb=\sb
\renewcommand\sp[1]{\ifmmode
	   \oldmsp{#1}%
	 \else\strut\raise.85ex\hbox{\scriptsize #1}\fi}
\renewcommand\sb[1]{\ifmmode
	   \oldmsb{#1}%
	 \else\strut\raise-.54ex\hbox{\scriptsize #1}\fi}
\newcommand\msp[1]{\ifmmode
	   \oldmsp{#1}
	 \else \math{\oldmsp{#1}}\fi}
\newcommand\msb[1]{\ifmmode
	   \oldmsb{#1}
	 \else \math{\oldmsb{#1}}\fi}
\newcommand\supon{\catcode`\^=7}
\newcommand\supoff{\catcode`\^=12}
\newcommand\subon{\catcode`\_=8}
\newcommand\suboff{\catcode`\_=12}
\newcommand\supsubon{\supon \subon}
\newcommand\supsuboff{\supoff \suboff}
}

\newcommand\actcharon{\catcode`\~=13}
\newcommand\actcharoff{\catcode`\~=12}

\newcommand\paramon{\catcode`\#=6}
\newcommand\paramoff{\catcode`\#=12}

\comment{And now to turn us totally on and off...}

\newcommand\reservedcharson{\commenton \mathshifton \atabon \supsubon \actcharon
	\paramon}

\newcommand\reservedcharsoff{\commentoff \mathshiftoff \ataboff
	\supsuboff \actcharoff \paramoff}

\newcommand\nojoe[1]{\reservedcharson#1\reservedcharsoff}

\catcode`@=12
\reservedcharsoff

\reservedcharson

\comment{ Must have ONLY ONE of these... trust these macros, they work

}
\reservedcharsoff

\actcharon
\if\submitms y
\newcommand\widedeltab{deluxetable}
\else
\newcommand\widedeltab{deluxetable*}
\received{2010 May 5}
\revised{2011 January 31}
\revised{2011 August 13}
\accepted{}
\fi

\if\submitms y
\slugcomment{\tablebox{
Submitted to {\em ApJ} 2011 January 31.}}
\else
\slugcomment{Submitted to {\em ApJ}. 
\comment{Revision DRAFT of {\today} \now.}}
\journalinfo{}
\fi

\begin{document}

\title{\textit{Spitzer} Secondary Eclipses of WASP-18b}

\author{Sarah Nymeyer\altaffilmark{1}}
\author{Joseph Harrington\altaffilmark{1}}
\author{Ryan A.\ Hardy\altaffilmark{1}}
\author{Kevin B.\ Stevenson\altaffilmark{1}}
\author{Christopher J.\ Campo\altaffilmark{1}}
\author{Nikku Madhusudhan\altaffilmark{2}}
\author{Andrew Collier-Cameron\altaffilmark{3}}
\author{Thomas J. Loredo\altaffilmark{4}}
\author{Jasmina Blecic\altaffilmark{1}}
\author{William C.\ Bowman\altaffilmark{1}}
\author{Christopher B.\ T.\ Britt\altaffilmark{1}}
\author{Patricio Cubillos\altaffilmark{1}}
\author{Coel Hellier\altaffilmark{5}}
\author{Michael Gillon\altaffilmark{6}}
\author{Pierre F. L. Maxted\altaffilmark{3}}
\author{Leslie Hebb\altaffilmark{7}}
\author{Peter J. Wheatley\altaffilmark{8}}
\author{Don Pollacco\altaffilmark{9}}
\author{David R. Anderson\altaffilmark{5}}
\email{sarah.nymeyer@gmail.com}

\affil{\sp1 Planetary Sciences Group, Department of Physics,
  University of Central Florida, Orlando, FL 32816-2385, USA}
\affil{\sp2 Department of Astrophysical Sciences, Princeton
  University, Princeton, NJ 08544, USA}
\affil{\sp3 School of Physics and Astronomy, University of
  St. Andrews, North Haugh, Fife KY16 9SS, UK}
\affil{\sp4 Center for Radiophysics and Space Research, Space Sciences
  Building, Cornell University, Ithaca, NY 14853-6801, USA}  
\affil{\sp5 Astrophysics Group, Keele University, Staffordshire ST5
  5BG, UK}
\affil{\sp6 Institut d'Astrophysique et de G\'eophysique, Universit\'e
  de Li\`ege, All\'ee du 6 Ao\^ut 17, Bat. B5C, 4000 Li\`ege, Belgium}
\affil{\sp7 Department of Physics and Astronomy, Vanderbilt
  University, Nashville, TN 37235, USA}
\affil{\sp8 Department of Physics, University of Warwick, Coventry,
  CV4 7AL, UK}
\affil{\sp9 Astrophysics Research Centre, School of Mathematics &
  Physics, Queen's University, University Road, Belfast, BT7 1NN, UK}

\begin{abstract}

The transiting exoplanet WASP-18b was discovered in 2008 by the Wide
Angle Search for Planets (WASP) project.  The \textit{Spitzer}
Exoplanet Target of Opportunity Program observed secondary eclipses of
WASP-18b using \textit{Spitzer}'s Infrared Array Camera (IRAC) in the
3.6 {\micron} and 5.8 {\micron} bands on 2008 December 20, and in the
4.5 {\micron} and 8.0 {\micron} bands on 2008 December 24.  We report
eclipse depths of \math{0.30\pm{0.02}\%, 0.39\pm{0.02}\%,
  0.37\pm{0.03}\%, 0.41\pm{0.02}\%}, and brightness temperatures of
3100\pm{90}, 3310\pm{130}, 3080\pm{140} and 3120\pm{110} K in order of
increasing wavelength. WASP-18b is one of the hottest planets yet
discovered - as hot as an M-class star.  The planet's
pressure-temperature profile most likely features a thermal inversion.
The observations also require WASP-18b to have near-zero albedo and
almost no redistribution of energy from the day-side to the night side
of the planet.

\comment{\if\submitms y
\else
\hfill\herenote{DRAFT of {\today} \now}.
\fi
}
\end{abstract}

\section{INTRODUCTION}
\label{intro}

\if\submitms y
\clearpage
\fi
\newcommand\s{@{\hspace{5pt}}}
\atabon\begin{\widedeltab}{l\s c\s c\s c|l\s c\s c|l\s c\s c|l\s c\s c}
\if\submitms y
\tabletypesize{\tiny}
\fi
\tablecaption{\label{tab:run} Run Parameters}
\tablewidth{0pt}
\tablehead{
\multicolumn{2}{c}{} &
\multicolumn{2}{c}{\math{\lambda}}   &
\multicolumn{3}{c}{Pre-Observation}  &
\multicolumn{3}{c}{Main Observation} &
\multicolumn{3}{c}{Post-Observation} \\
\colhead{Obs.} &
\colhead{Date} &
\multicolumn{2}{c}{({\microns})} &
\colhead{Start\tablenotemark{a}} &
\colhead{Frame Time (s)} &
\colhead{Frames} &
\colhead{Start\tablenotemark{a}} &
\colhead{Frame Time (s)} &
\colhead{Frames} &
\colhead{Start\tablenotemark{a}} &
\colhead{Frame Time (s)} &
\colhead{Frames} 
}
\startdata
CH13    & 2008-12-20      & 3.6 & 5.8     & 820.57796     & 2, 12  & 260    & 820.61869     &   2, 12     & 1148  & 820.79721  & 2, 12   & 10   \\
CH24    & 2008-12-24      & 4.5 & 8.0     & 824.35796     & 12     & 185    & 824.38431     &   2, 12     & 1148  & 824.56278  & 2, 12   & 10
\enddata
\tablenotetext{a}{BJD - 2,454,000 ephemeris time}
\end{\widedeltab}\ataboff
\if\submitms y
\clearpage
\fi
\placetable{tab:run}

Among the more than 500 extrasolar planets discovered to
date\footnote[1]{For an up-to-date listing, see
  http://www.exoplanet.eu}, the over 100 close-orbiting gas giants
that transit their host stars have provided the most valuable clues to
their physical natures. The geometry of the transit gives a direct
measurement of the density of the host star and the surface gravity of
the planet
\citep{SeagerEtal2003apjPlanetStarParameters,SouthworthEtal2007mnrasHD209458b}. These
measurements can be combined with an estimate of the star's mass and
radius to provide estimates of the planet's mass, radius and
density. The closest-orbiting planets attain dayside temperatures
high enough to give observable secondary eclipses at thermal-infrared
wavelengths as they pass behind their host stars (e.g.,
\citealp{CharbonneauEtal2005apjTrES1,DemingEtal2005natHD209458b,SingEtal2009aaGroundbased}). 

Observations with the \textit{Spitzer Space Telescope} have revealed
that transiting gas giant planets can be divided into two classes
based on their infrared spectral energy distributions.  A subset of the
very hottest planets, with dayside temperatures in excess of 2000 K,
display molecular features of CO and H\sb{2}O in emission rather than
absorption, indicating the presence of a temperature increase with
height in the planet's photospheric layers.  Such temperature
inversions have been inferred from the flux ratios between the four
bandpasses of the Infrared Array Camera (IRAC,
\citealp{FazioEtal2004apjsIRAC}) aboard \textit{Spitzer}, centered at
3.6, 4.5, 5.8, and 8.0 {\micron} (channels 1--4), in the hot planets
HD209458b \citep{BurrowsEtal2008apjHD209458b,
  KnutsonEtal2008apjHD209458b}, XO-1b \citep{MachalekEtal2008XO-1b},
TrES-2b \citep{ODonovanEtal2010apjTres-2b}, TrES-4b
\citep{KnutsonEtal2009apjTres4}, XO-2b \citep{MachalekEtal2009XO-2b}
and WASP-1b \citep{WheatleyEtal2010}, among others.  This phenomenon
has been attributed to the presence of strongly-absorbing species such
as TiO and VO remaining in the gas phase in the upper atmosphere,
leading to the formation of a stratospheric temperature inversion
\citep{BurrowsEtal2008apjHD209458b, FortneyEtal2008TiOVO}. This led
\citeauthor{FortneyEtal2008TiOVO} to propose a scheme in which planets
are assigned to class pM (with hot stratospheres) or pL (without)
according to temperature, by analogy with the stellar M and L spectral
classes.  \citet{KnutsonEtal2010apjStellarActivity} suggest instead
that some planets lack inversions because the TiO and VO are destroyed
by UV radiation, while \citet{SpiegelEtal2009apjTiO} claimed that
TiO/VO may not be sufficiently abundant in the upper atmospheres to
produce the required
inversions.  \citet{ZahnleEtal2009apjphotochemistry} proposed additional
absorbers, like sulphur species, which could potentially form thermal
inversions.

WASP-18b is a prime candidate for secondary-eclipse observations to
test for the presence of a hot stratosphere
\citep{HellierEtal2009natWASP18b}. It orbits an F6-type star with a
period of just 0.94 days and is expected to attain a blackbody
equilibrium temperature approaching 2400 K, assuming zero albedo and
efficient transport of heat from the dayside to the nightside of the
planet. With less efficient transport, an even higher dayside
temperature is expected. The planet is unusual because of its very
high mass, \math{M\sb{\rm p}=10.43\pm{0.30}} Jupiter masses,
\math{M\sb{\rm Jup}}, and modest radius, \math{R\sb{\rm
    p}=1.165\pm{0.055}} Jupiter radii, \math{R\sb {\rm Jup}}
\citep{SouthworthEtal2009apjWASP18b}, which give it a surface gravity
an order of magnitude greater than that of any known transiting planet
likely to belong to the pM class. The pressure scale height in its
photospheric layers should therefore be an order of magnitude smaller
than for planets of similar temperature, providing an important new
dimension for tests of planetary atmospheric models in the presence of
strong external irradiation. The small but precisely known orbital
eccentricity \math{e}=0.0085 \math{\pm0.0008}
\citep{Triaudetal2010apjSpinOrbitMeas} may impart slightly
faster-than-synchronous rotation, which could help to redistribute
heat from the dayside to the nightside of the planet.

WASP-18b was therefore selected as a candidate for observation as part
of our \textit{Spitzer} Exoplanet Target of Opportunity Program
shortly after its discovery was confirmed. In Sections 2 and 3 of this
paper we describe the observations and the analysis of the data. In
Section 4 we discuss the constraints imposed on the thermal structure
of the planetary atmosphere by the IRAC fluxes, in Section 5 we
compare the eccentricity and orientation of the orbit from the timing
and duration of the secondary eclipse with the values derived from
radial-velocity observations, and in Section 6 we present our
conclusions. 

\section{OBSERVATIONS}
\label{sec:obs}

\textit{Spitzer}'s IRAC instrument observed (program 50517) two
secondary eclipses (see Table \ref{tab:run}). After each observation,
an offset 6-minute full-array sequence confirmed the lack of
persistent bad pixels near the stellar position. There are two
independent analyses of these data, one presented here and one by
\citet{MachalekEtal2010}. 

IRAC exhibits some sources of systematic noise that must be taken into
account when planning and analyzing observations. A positional
sensitivity exists in the 3.6 and 4.5 {\micron} channels
\citep{CharbonneauEtal2005apjTrES1,KnutsonEtal2009apjHD149026bphase,
 MachalekEtal2010,StevensonEtal2010natGJ436b}, and a time-varying
sensitivity (''ramp'') exists in the 5.8 and 8.0 {\micron} channels
\citep{CharbonneauEtal2005apjTrES1,HarringtonEtal2007natHD149026b}.

\begin{figure}[t]
\if\submitms y
  \setcounter{fignum}{\value{figure}}
  \addtocounter{fignum}{1}
  \newcommand\fignam{f\arabic{fignum}.eps}
\else
  \newcommand\fignam{wa018bs41_preflash.eps}
\fi
\includegraphics[width=\columnwidth, clip]{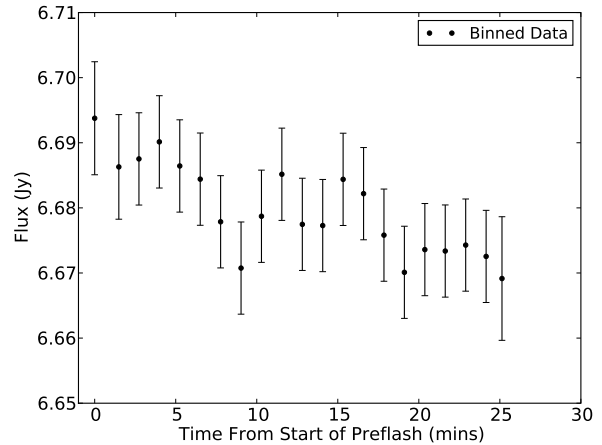}
\figcaption{\label{fig:preflash}
Preflash light curve.  These are 8 {\micron} data, analyzed with
aperture photometry.  The preflash source is bright compared to
WASP-18, which allows the array sensitivity to stabilize before the
science observations.  Without a preflash, similar observations
generally show a steeper and longer ramp in the eclipse observations.
This may be the first descending preflash, attributed to
an even brighter source in the immediately preceeding program.
}
\end{figure}

The time-varying sensitivity in the 8.0 {\micron} channel manifests as
an apparent increase in flux with time.  The rate of increase depends
on the number of photons received by each pixel, and is believed to be
caused by charge trapping.  The effect is successfully reduced by
staring at a bright diffuse source, in this instance an HII ionized
region, prior to the main observation (a "preflash").  The large
number of photons quickly saturates the detector, resulting in a
smaller rate of increase in the main observation than is seen without
a preflash \citep{KnutsonEtal2009apjHD149026bphase}.  We observed a
30-minute preflash prior to the December 24 event (see Figure
\ref{fig:preflash}), which exhibits a decreasing ramp, unlike previous
observations
\citep{KnutsonEtal2009apjHD149026bphase,CampoEtal2010apjWASP18b}.
This is attributable to the previous IRAC observation of the bright
extended source IC1396a.

The 5.8 {\micron} channel exhibits a decreasing ramp, so
\textit{Spitzer} stared at the target for 62 minutes prior to the
observation to allow time for the detector to stabilize.  In order to
minimize the positional sensitivity at 3.6 and 4.5 {\micron}, each
observation used fixed pointing.

\if\submitms y
\clearpage
\fi
\if\submitms y
\clearpage
\fi

\if\submitms y
\clearpage
\fi
\atabon\begin{\widedeltab}{lr@{\,\pm\,}lr@{\,\pm\,}lr@{\,\pm\,}lr@{\,\pm\,}l}
\if\submitms y
\tabletypesize{\scriptsize}
\fi
\tablecaption{\label{tab:fits} Details of the Analysis and Results From Light Curve Fit}
\tablewidth{0pt}
\tablehead{
\colhead{Parameter} &
\mctc{3.6 {\microns}} &
\mctc{4.5 {\microns}} &
\mctc{5.8 {\microns}} &
\mctc{8.0 {\microns}}
}
\startdata
Array Position (\math{\bar{x}}, pix)                             &        \mctc{30.24}        &         \mctc{23.20}         &        \mctc{24.41}        &      \mctc{24.17}         \\
Array Position (\math{\bar{y}}, pix)                             &        \mctc{23.89}        &         \mctc{24.68}         &        \mctc{24.05}        &      \mctc{22.32}         \\
Position Consistency\tablenotemark{a} (\math{\delta\sb{x}}, pix) &        \mctc{0.018}        &        \mctc{0.019}          &       \mctc{0.028}         &      \mctc{0.022}         \\
Position Consistency\tablenotemark{a} (\math{\delta\sb{y}}, pix) &        \mctc{0.018}        &        \mctc{0.038}          &       \mctc{0.015}         &      \mctc{0.021}         \\
Aperture Size (pix)                                              &        \mctc{3.50}         &        \mctc{3.00}           &       \mctc{3.75}          &      \mctc{3.50}          \\
Sky Annulus Inner Radius (pix)                                   &         \mctc{7}           &         \mctc{7}             &        \mctc{7}            &       \mctc{7}            \\
Sky Annulus Outer Radius (pix)                                   &         \mctc{12}          &        \mctc{12}             &       \mctc{12}            &      \mctc{12}            \\
System Flux (\math{F\sb{\rm s}}, {\micro}Jy)                     &       168080 & 140         &        104300 & 600          &        69690 & 20          &      37450 & 10           \\
Eclipse Depth (\math{F\sb{\rm p}/F\sb{\rm s}}, 
Brightness Temperature (K)                                       &         3100 &  90         &          3310 & 130          &         3080 & 140         &       3120 & 110          \\
Eclipse Mid-time (\math{t\sb{\rm mid}}, phase)\tablenotemark{b}  &       0.4995 & 0.0007      &        0.4985 & 0.0006       &       0.4995 & 0.0007      &     0.4985 & 0.0006       \\
Eclipse Mid-time (\math{t\sb{\rm mid}}, BJD - 2,454,000)\tablenotemark{b} & 820.7160 & 0.0006 &      824.4809 & 0.0005       &     820.7160 & 0.0006      &   824.4809 & 0.0005       \\
Eclipse Duration (\math{t\sb{\rm 4-1}}, sec)\tablenotemark{b}    &         8010 &  60         &          8010 &  60          &         8010 & 60          &       8010 &  60          \\
Ingress (\math{t\sb{\rm 2-1}}) and Egress (\math{t\sb{\rm 4-3}}) Times (sec)\tablenotemark{b} &  \mctc{857} &  \mctc{857}    &        \mctc{857}          &      \mctc{857}          \\
Ramp Name                                                        &        \mctc{Linear}       &        \mctc{Linear}         & \mctc{Falling Exponential} & \mctc{Rising Exponential} \\
Ramp, Linear Term (\math{r\sb{1}})                               &        0.005 &  0.001      &  \math{-0.006} & 0.003       &         \mctc{0}           &          \mctc{0}         \\
Ramp, Curvature (\math{r\sb{2}})                                 &          \mctc{0}          &           \mctc{0}           &           14 & 1           &         17 & 1            \\
Ramp, Time Offset (\math{r\sb{3}})                               &          \mctc{0}          &           \mctc{0}           &       \mctc{-0.035689}     &      \mctc{0.082618}      \\
Intrapixel, Quadratic Term in x (\math{p\sb{2}})                &          \mctc{0}          &          0.47 & 0.13         &         \mctc{0}           &         \mctc{0}          \\
Intrapixel, Cross Term (\math{p\sb{3}})                         &          \mctc{0}          &  \math{-0.12} & 0.01         &         \mctc{0}           &         \mctc{0}          \\
Intrapixel, Linear Term in y (\math{p\sb{4}})                   &        0.067 & 0.004       &          \mctc{0}            &         \mctc{0}           &         \mctc{0}          \\
Intrapixel, Linear Term in x (\math{p\sb{5}})                   &  \math{-0.086} & 0.003     &  \math{-0.33} & 0.06         &         \mctc{0}           &         \mctc{0}          \\
Total frames                                                     &           \mctc{1148}      &          \mctc{1148}         &         \mctc{1148}        &        \mctc{1148}        \\
Good frames                                                      &           \mctc{1142}      &           \mctc{987}         &          \mctc{996}        &        \mctc{1031}        \\
Rejected frames ({\%})                                           &           \mctc{0}         &           \mctc{14}          &          \mctc{13}         &         \mctc{10}         \\
Standard Deviation of Normalized Residuals                       &      \mctc{0.002428}       &       \mctc{0.003485}        &       \mctc{0.003738}      &         \mctc{0.002928}   \\ 
Uncertainty scaling factor                                       &      \mctc{0.31674}        &        \mctc{0.47164}        &       \mctc{0.57274}       &         \mctc{0.47792}    \
\enddata
\tablenotetext{a}{RMS frame-to-frame position difference}
\tablenotetext{b}{Duration and ingress/egress time are each a single
  parameter shared among all four wavelengths.  Eclipse mid-times are
  a single parameter for each pair of channels observed together.} 
\end{\widedeltab}\ataboff
\if\submitms y
\clearpage
\fi
\placetable{tab:fits}

\section{DATA ANALYSIS}
\label{sec:anal}

\if\submitms y
\clearpage
\fi
\begin{figure*}[t]
\if\submitms y
  \setcounter{fignum}{\value{figure}}
  \addtocounter{fignum}{1}
  \newcommand\fignama{f\arabic{fignum}a.ps}
  \newcommand\fignamb{f\arabic{fignum}b.ps}
  \newcommand\fignamc{f\arabic{fignum}c.ps}
\else
  \newcommand\fignama{wasp18-raw.ps}
  \newcommand\fignamb{wasp18-bin.ps}
  \newcommand\fignamc{wasp18-norm.ps}
\fi
\strut\hfill
\includegraphics[width=0.32\textwidth, clip]{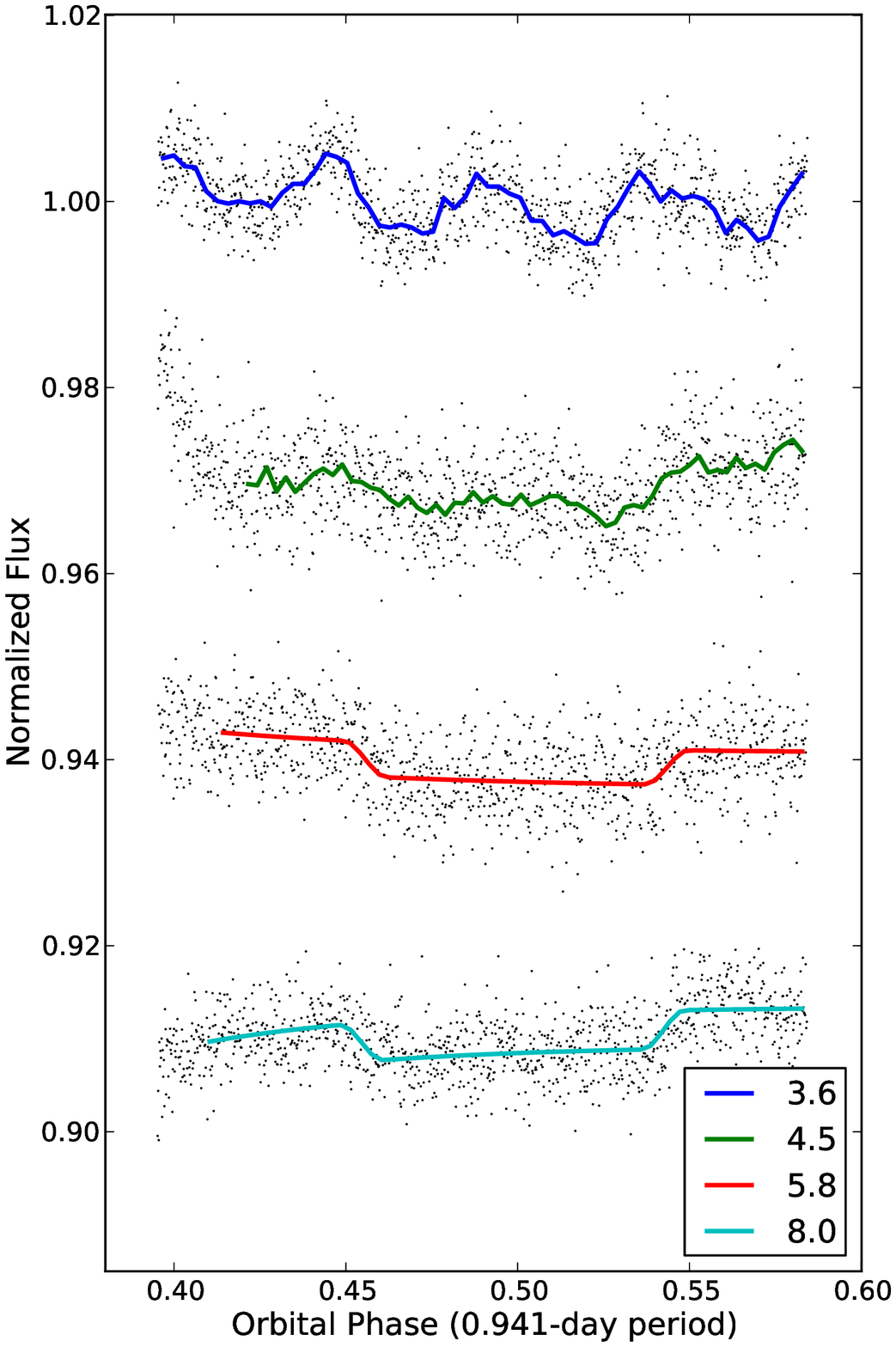}
\includegraphics[width=0.32\textwidth, clip]{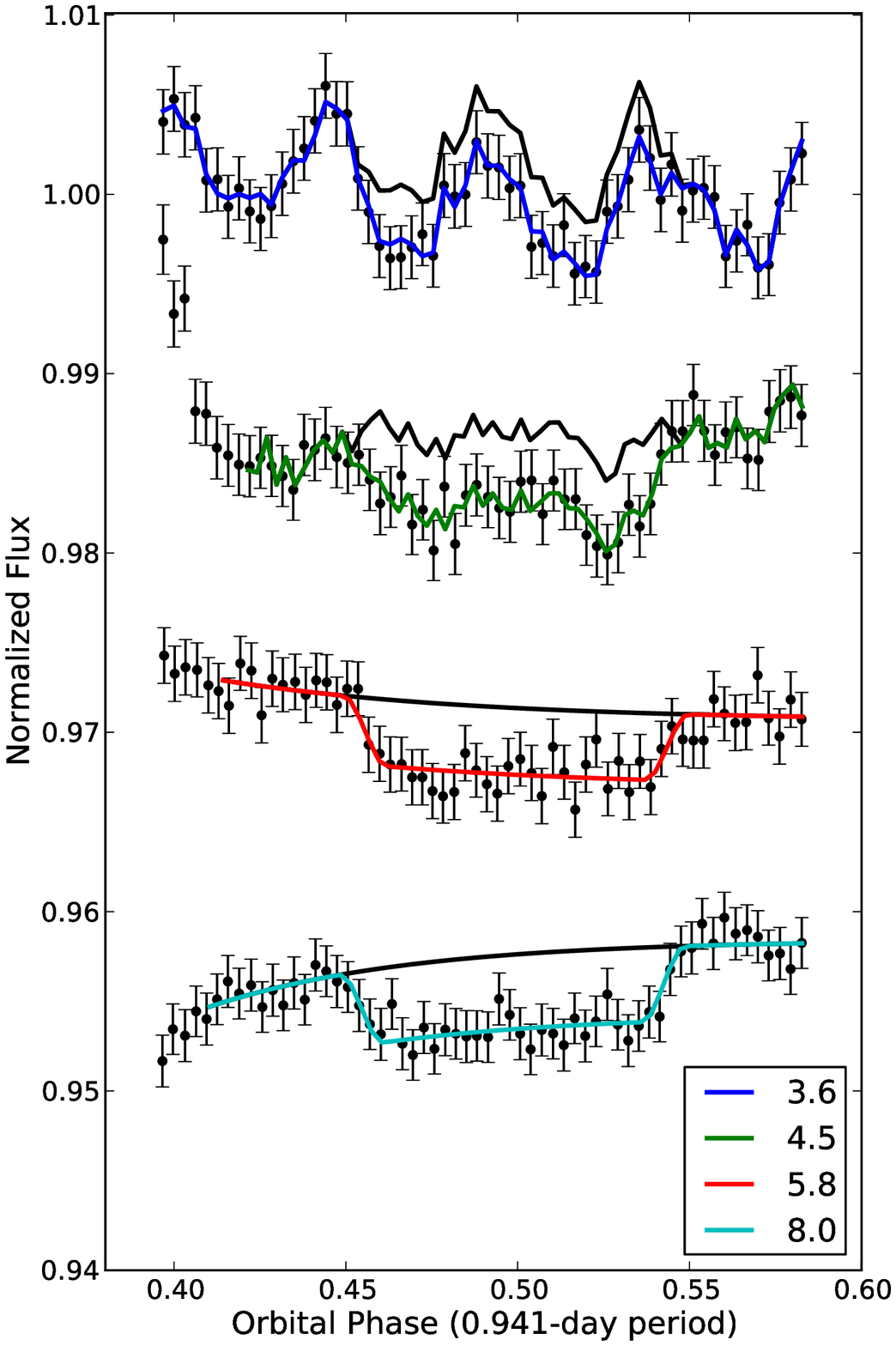}
\includegraphics[width=0.32\textwidth, clip]{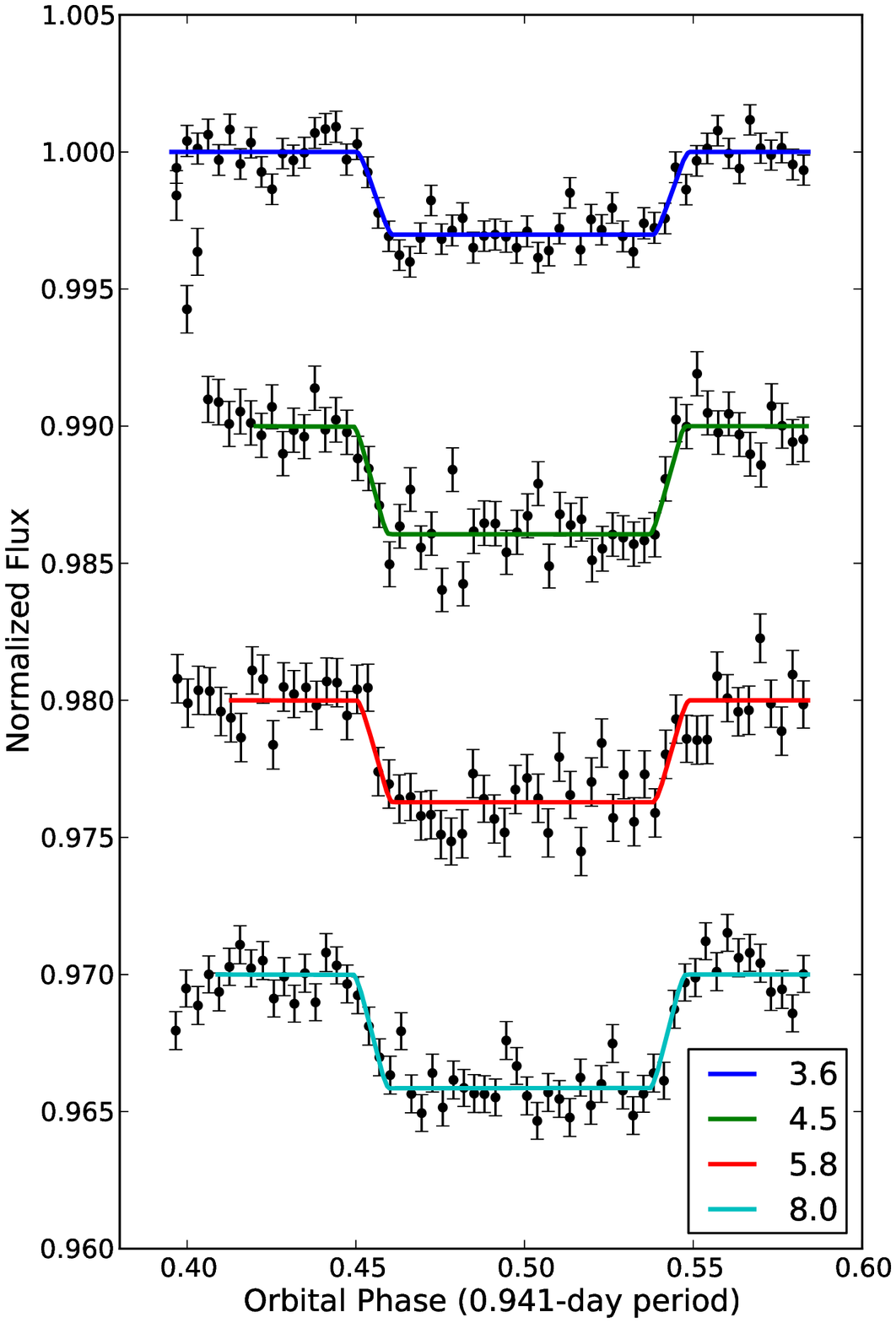}
\hfill\strut
\figcaption{\label{fig:lightcurves}
Raw (left), binned (center), and systematics-corrected (right)
secondary-eclipse light curves of WASP-18b in the four IRAC channels,
normalized to the mean system flux within the fitted data.  Colored
lines are the best-fit models; black curves omit their eclipse model
elements. A few initial points in all channels are not fit, as
indicated, to allow the telescope and instrument to stabilize.
}
\end{figure*}
\if\submitms y
\clearpage
\fi

All data are \textit{Spitzer} Basic Calibrated Data (BCD) frames using
version S18.7.0 of \textit{Spitzer}'s preprocessing pipeline.  This
pipeline removes all well-understood instrumental signatures and
produces a flux-calibrated image \citep{FazioEtal2004apjsIRAC}.  We
first account for light travel time in the solar system by converting
to Barycentric Julian Date (BJD), and then mask all pixels in the
\textit{Spitzer}-supplied permanently bad pixel mask.  We find the
remaining bad pixels by grouping sets of 64 frames and doing a
two-iteration outlier rejection at each pixel location.  Within each
array position in each set, this routine calculates the standard
deviation from the \textit{median}, masks any pixels with greater than
\math{4\sigma} deviation, and repeats this procedure once.  Masked
pixels do not participate in the analysis.

WASP-18 is very bright relative to the background.  A 2D Gaussian fit
to data within 4 pixels of the stellar brightness peak determines the
stellar center in each frame.  The light curve comes from
5{\by}-interpolated aperture photometry
\citep{HarringtonEtal2007natHD149026b}, excluding frames with masked
pixels in the photometry aperture and not using masked pixels in sky
level averages.  Table \ref{tab:fits} presents photometry parameters.
We vary the aperture radius between 2.0 and 5.0 pixels in 0.25-pixel
increments, choosing the one with the best light-curve fit as
described below.

We model the intrapixel variation affecting the 3.6 and 4.5 {\micron}
channels with a second-order, two-dimensional polynomial
\citep{KnutsonEtal2008apjHD209458b,StevensonEtal2010natGJ436b,CampoEtal2010apjWASP18b},
\begin{eqnarray}
\label{eq:vip} V\sb{\rm IP}(x,y) = p\sb1y\sp2 + p\sb2x\sp2 + p\sb3xy +
p\sb4y + p\sb5x + 1,
\end{eqnarray} 
\noindent where \math{x} and \math{y} are the centroid coordinates
relative to the pixel center nearest the median position, and
\math{p\sb{1}} - \math{p\sb{5}} are free parameters.  The systematics
had little to no dependence on the quadratic \math{y} term, so
\math{p\sb{1}} is fixed to zero for all models.  The ramps for the 3.6
and 4.5 {\micron} channels use a linear model,
  \begin{eqnarray}
\label{eq:linramp}
R(t) = r\sb{1}(t - 0.5) + 1,
\end{eqnarray}
\noindent where \math{t} is orbital phase and \math{r\sb{1}} is a free
parameter.  We model the 5.8 {\micron} channel with a falling exponential
\begin{eqnarray}
\label{eq:fallexp}
R(t) = 1 + \exp \left( -r\sb{2}[t - r\sb{3}] \right),
\end{eqnarray}
\noindent and the 8.0 {\micron} channel with a rising exponential
\citep{HarringtonEtal2007natHD149026b}  
\begin{eqnarray}
\label{eq:rexp}
R(t) = 1 - \exp \left( -r\sb{2}[t - r\sb{3}] \right),
\end{eqnarray}
\noindent where \math{r\sb{2}} and \math{r\sb{3}} are free
parameters. The \math{r\sb{3}} term is fixed to its best-fit value.
The eclipse, \math{E(t)}, is a \citet{MandelAgol2002apjLightcurves}
model which includes the time of secondary eclipse, \math{t\sb{1}} to
\math{t\sb{4}} duration (1\sp{st} to 4\sp{th} contact), ingress/egress
time, and eclipse depth.

The single-channel light curve model is 
\begin{eqnarray}
\label{eq:lcmodel}
F(x, y, t) = F\sb{\rm s}V\sb{\rm IP}(x, y)R(t)E(t),
\end{eqnarray}
\noindent where \math{F(x, y, t)} is the flux measured from
interpolated aperture photometry and \math{F\sb{s}} is the (constant)
system flux outside of eclipse, including the planet.  We dropped a
small number of initial frames in each light curve (0, 150, 100, and
80 frames in order of increasing wavelength) to allow the pointing and
instrument to stabilize.  The model lines in Figure
\ref{fig:lightcurves} show which points are included, as do the
electronic light curve files.  We fit Eq.\ \ref{eq:lcmodel} to the
data using a least-squares minimizer.  Because the Spitzer pipeline
usually overestimates uncertainties, we re-scale the photometric
uncertainties to produce a reduced \math{\chi\sp{2}} of 1 and re-run
the fit.  This typically converges in one iteration.  For a given
photometric set the scaling factors are almost identical for all
models, so we choose one for use with all models.

Because the underlying physics of the systematics have not been
characterized sufficiently to find an expression that fits well in
every case, it is possible that, for some \textit{Spitzer} data sets,
different investigators will find different values for key parameters
when using different systematic models and photometry parameters. This
occurred, for example, when \citet{KnutsonEtal2009apjHD149026bphase}
re-analyzed the data of \citet{HarringtonEtal2007natHD149026b},
finding an eclipse about half as deep.  The \math{\chi\sp{2}} minimum
that \citeauthor{KnutsonEtal2009apjHD149026bphase} found was present
in the correlation plots of
\citeauthor{HarringtonEtal2007natHD149026b}, but it was just a local
minimum for that model.  While models for deep eclipses, such as those
presented here and by \citet{CampoEtal2010apjWASP18b}, should
generally produce compatible results even with different systematic
models, weak eclipses such as those of
\citet{HarringtonEtal2007natHD149026b} and
\citet{StevensonEtal2010natGJ436b} are more dependent on the details
of fitting. 

All of the published results of our current pipeline \citep[and this
  work]{StevensonEtal2010natGJ436b, CampoEtal2010apjWASP18b} result
from testing a variety of models for each systematic, as well as
assessing the best photometry aperture and stabilization time for each
data set.  We test linear, quadratic, quartic-in-log-time, falling or
rising exponential, logarithmic-plus-quadratic, and
logarithmic-plus-linear ramps, and a variety of polynomial intrapixel
models, before choosing the final models.  Most of the possibilities
produce obvious bad fits.  For this paper we select the best two
models in each channel and fit them for all apertures.

For each channel, photometry using the various aperture sizes produces
slightly different data sets.  We must select first the best data set
and then the best model, but \math{\chi\sp{2}} and related fitting
criteria only compare different models to a single data set; they are
inappropriate for deciding between models fit to different data sets.
For data sets from different apertures, we choose the one with the
smallest standard deviation of normalized residuals (SDNR) with
respect to the system flux for a given model and repeat for each of
several models.  This generally results in a consistent choice of the
best aperture size among different models. 

\begin{figure}[t]
\if\submitms y
  \setcounter{fignum}{\value{figure}}
  \addtocounter{fignum}{1}
  \newcommand\fignam{f\arabic{fignum}.eps}
\else
  \newcommand\fignam{ch1sdnrdb.eps}
\fi
\includegraphics[width=\columnwidth, clip]{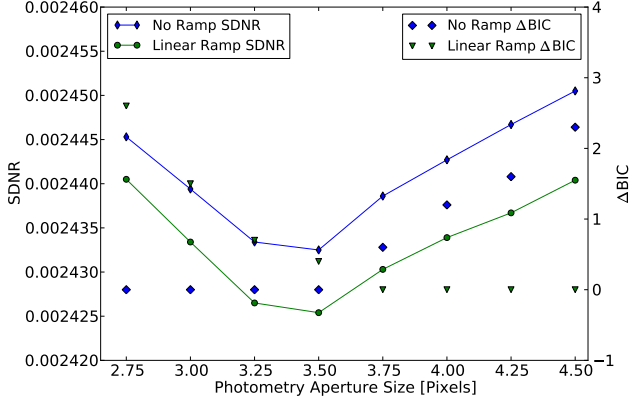}
\figcaption{\label{fig:ch1sdnr}
Channel 1. SDNR and \math{\Delta}BIC vs. aperture size.  A lower value indicates a better model fit.
}
\end{figure}

\begin{figure}[t]
\if\submitms y
  \setcounter{fignum}{\value{figure}}
  \addtocounter{fignum}{1}
  \newcommand\fignam{f\arabic{fignum}.eps}
\else
  \newcommand\fignam{ch2sdnrdb.eps}
\fi
\includegraphics[width=\columnwidth, clip]{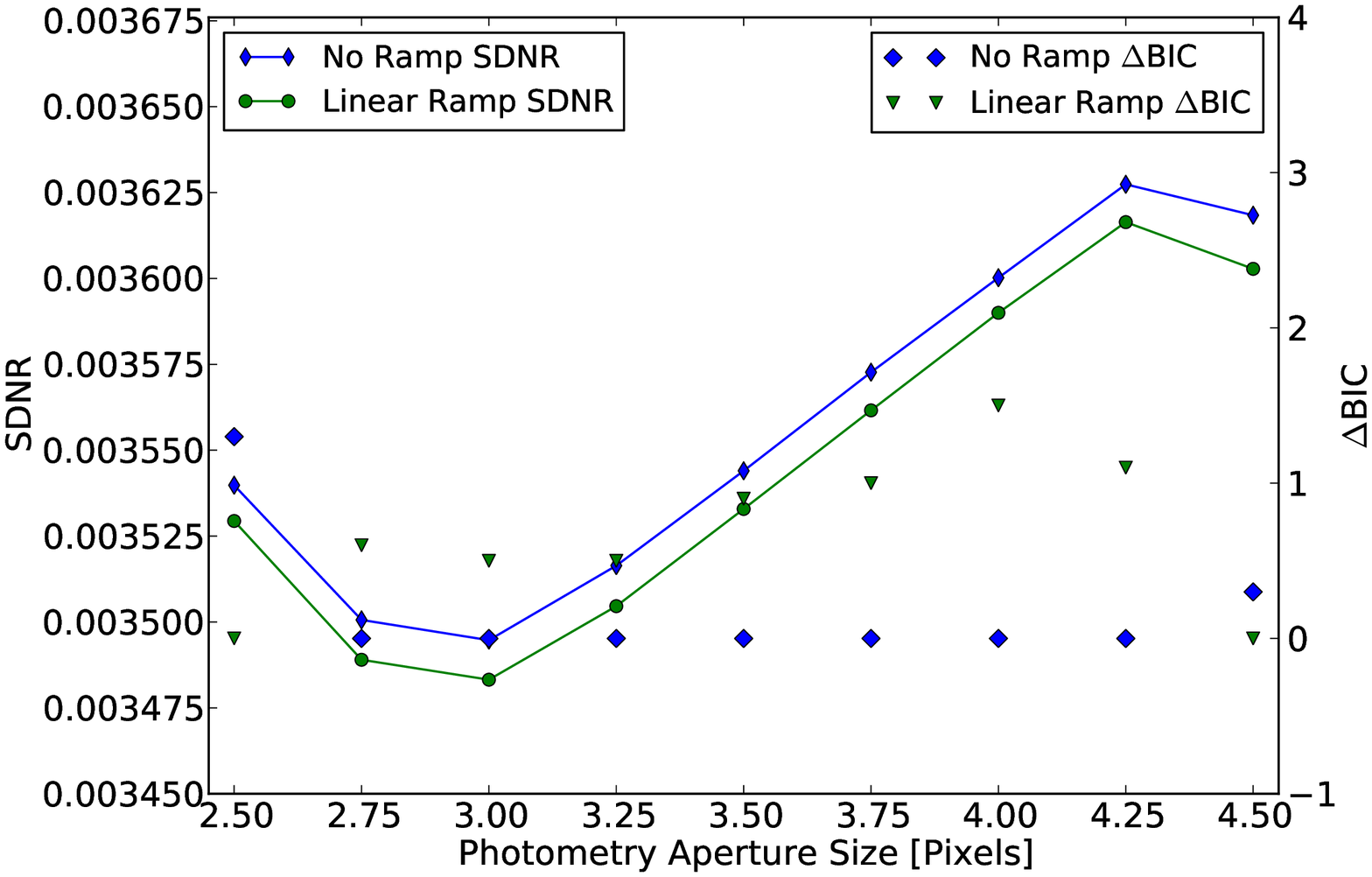}
\figcaption{\label{fig:ch2sdnr}
Channel 2. SDNR and \math{\Delta}BIC vs. aperture size.  A lower value indicates a better model fit.
}
\end{figure}

\begin{figure}[t]
\if\submitms y
  \setcounter{fignum}{\value{figure}}
  \addtocounter{fignum}{1}
  \newcommand\fignam{f\arabic{fignum}.eps}
\else
  \newcommand\fignam{ch3sdnrdb.eps}
\fi
\includegraphics[width=\columnwidth, clip]{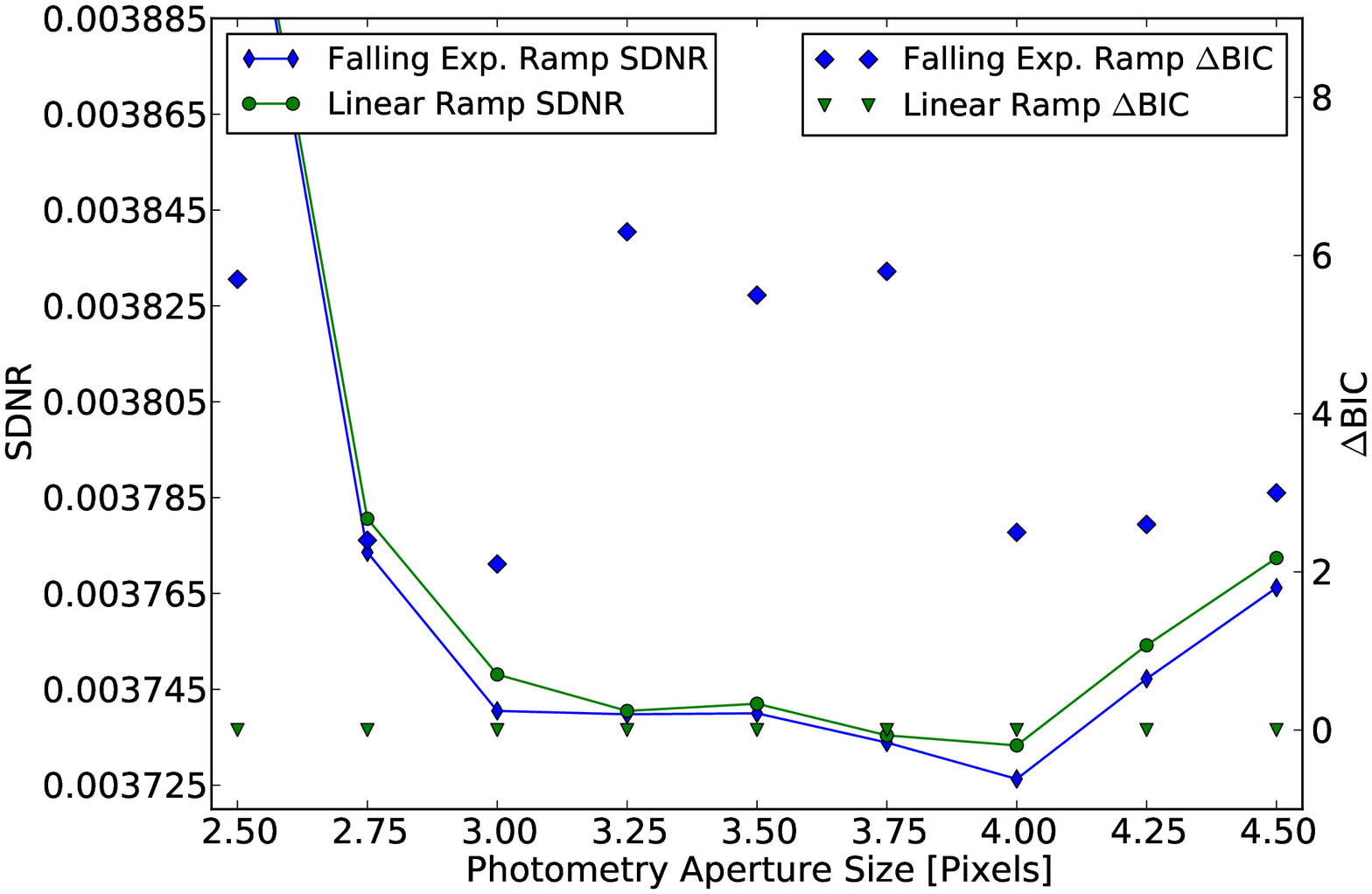}
\figcaption{\label{fig:ch3sdnr}
Channel 3. SDNR and \math{\Delta}BIC vs. aperture size.  A lower value indicates a better model fit.
}
\end{figure}

\begin{figure}[t]
\if\submitms y
  \setcounter{fignum}{\value{figure}}
  \addtocounter{fignum}{1}
  \newcommand\fignam{f\arabic{fignum}.eps}
\else
  \newcommand\fignam{ch4sdnrdb.eps}
\fi
\includegraphics[width=\columnwidth, clip]{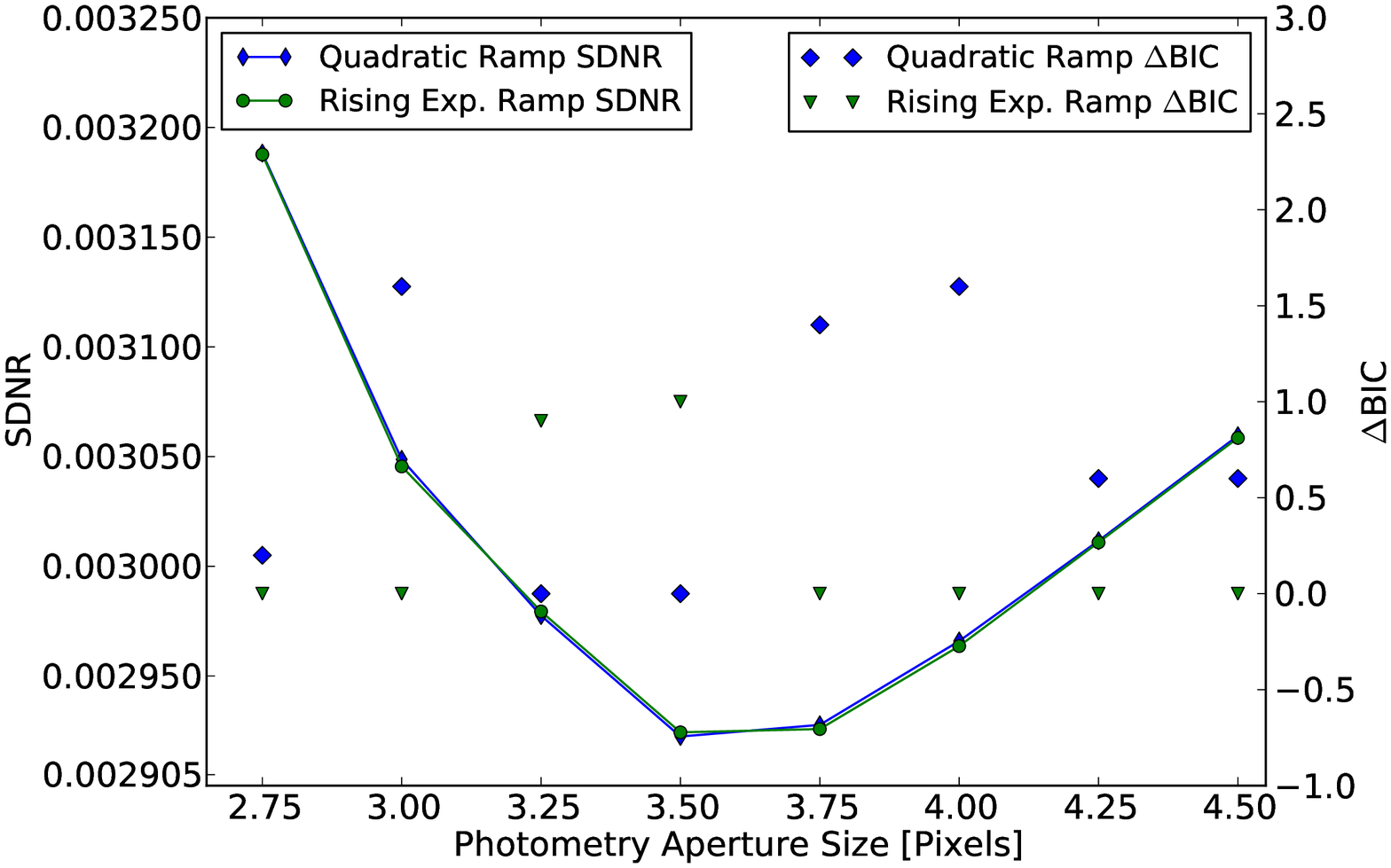}
\figcaption{\label{fig:ch4sdnr}
Channel 4. SDNR and \math{\Delta}BIC vs. aperture size.  A lower value indicates a better model fit.
}
\end{figure}

Once we have the optimal aperture size, we then compare the models.
Since adding additional parameters to a model will always produce a
better fit, we use fitting criteria that properly penalize the
addition of parameters.  As described by
\citet{CampoEtal2010apjWASP18b}, we apply both the Akaike Information
Criterion, 
\begin{equation}
{\rm AIC} = \chi\sp{2} + 2k,
\end{equation}
\noindent where \math{k} is the number of free parameters, and the
Bayesian Information Criterion, 
\begin{equation}
{\rm BIC} = \chi\sp{2} + k\ln N,
\end{equation}
\noindent where \math{N} is the number of data points
\citep{Liddle2007apjInfoCrit}. A lower information criterion value
indicates a better model.  Figures \ref{fig:ch1sdnr},
\ref{fig:ch2sdnr}, \ref{fig:ch3sdnr}, and \ref{fig:ch4sdnr} present
SNDR and BIC with the two main candidate models and aperture sizes for
each wavelength.  Our final joint model fit, with 28 free parameters,
combines the eclipse durations for all channels and pairs the
simultaneously observed mid-times.  It resulted in an AIC of 4176 and
a BIC of 4302.

To assess parameter uncertainties and correlations we explore phase
space with a Metropolis random-walk Markov-chain Monte Carlo (MCMC)
routine.  Each chain began at the least-squares minimum.  If any step
in the chain ever beat the minimum, it would indicate an even deeper
minimum at the bottom of the basin of attraction just entered, so the
routine would discard the MCMC data, re-run the minimizer, and
re-start the Markov chain.  The routine runs a ``burn-in'' of at least
\math{10\sp{5}} iterations to forget the starting conditions, and then
runs four million iterations. 

\begin{figure}[thb]
\if\submitms y
  \setcounter{fignum}{\value{figure}}
  \addtocounter{fignum}{1}
  \newcommand\fignama{f\arabic{fignum}a.eps}
  \newcommand\fignamb{f\arabic{fignum}b.eps}
  \newcommand\fignamc{f\arabic{fignum}c.eps}
  \newcommand\fignamd{f\arabic{fignum}d.eps}
\else
  \newcommand\fignama{ch1rms.eps}
  \newcommand\fignamb{ch2rms.eps}
  \newcommand\fignamc{ch3rms.eps}
  \newcommand\fignamd{ch4rms.eps}
\fi
\includegraphics[width=0.22\textwidth, clip]{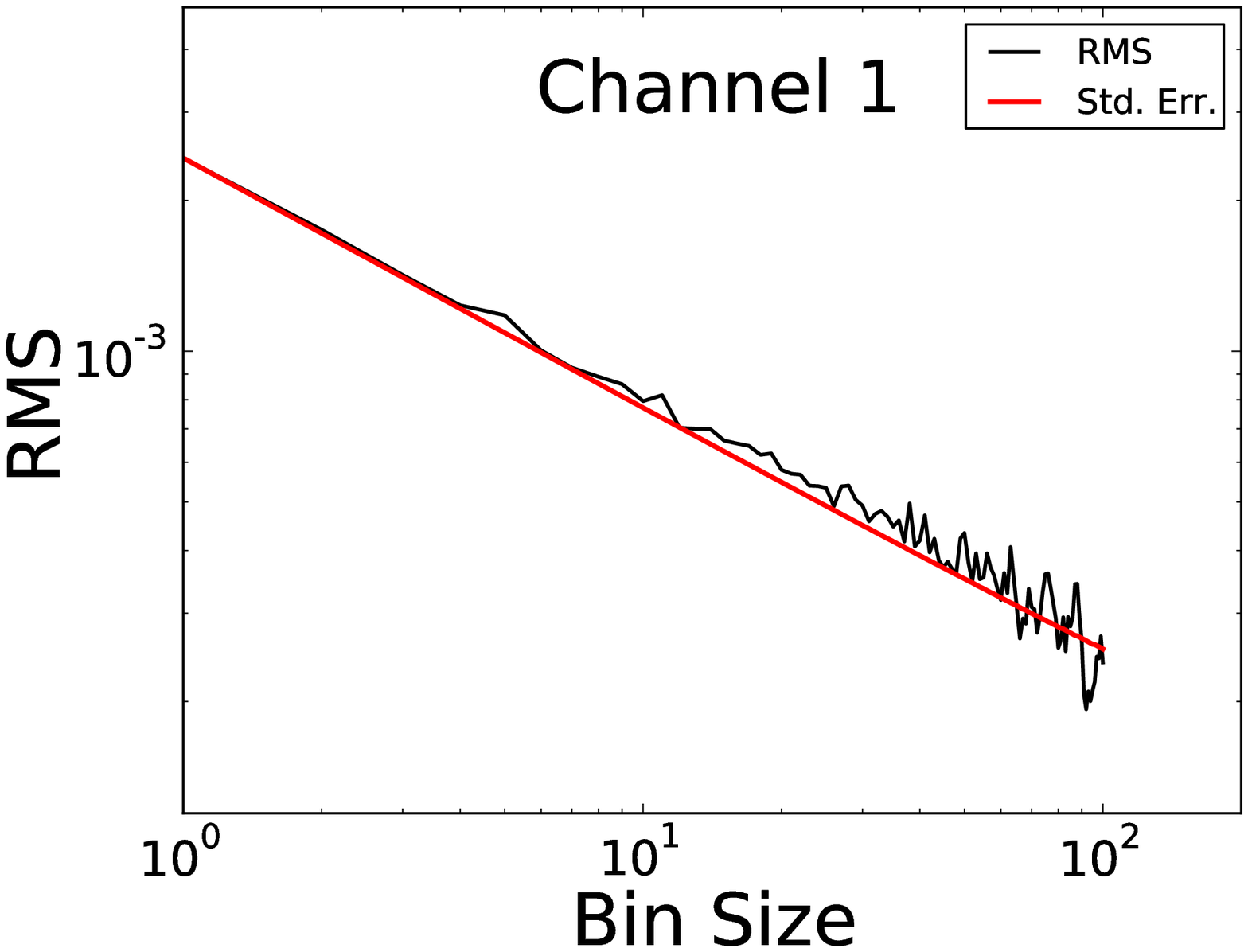}
\hfill
\includegraphics[width=0.22\textwidth, clip]{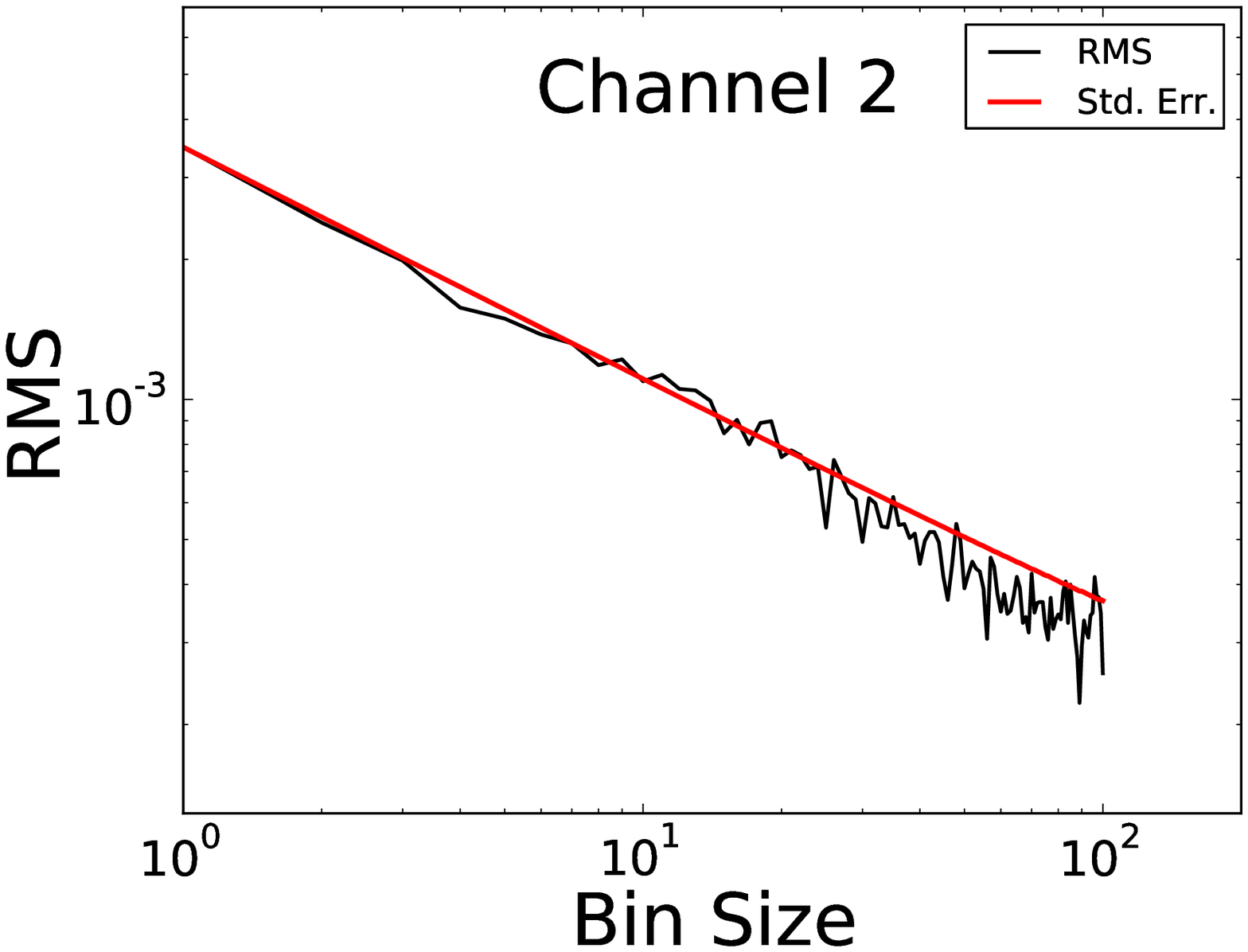}
\newline
\strut\newline
\includegraphics[width=0.22\textwidth, clip]{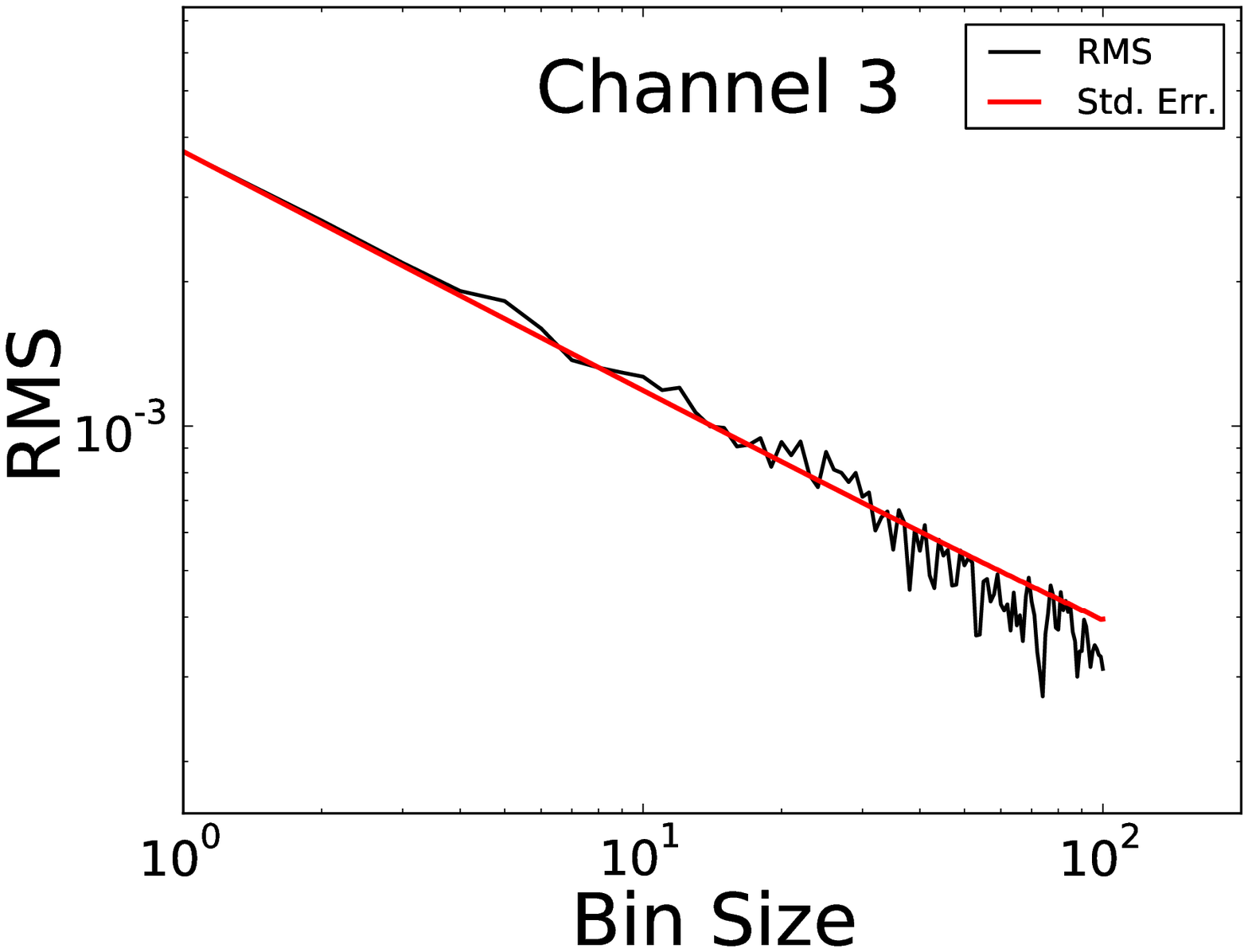}
\hfill
\includegraphics[width=0.22\textwidth, clip]{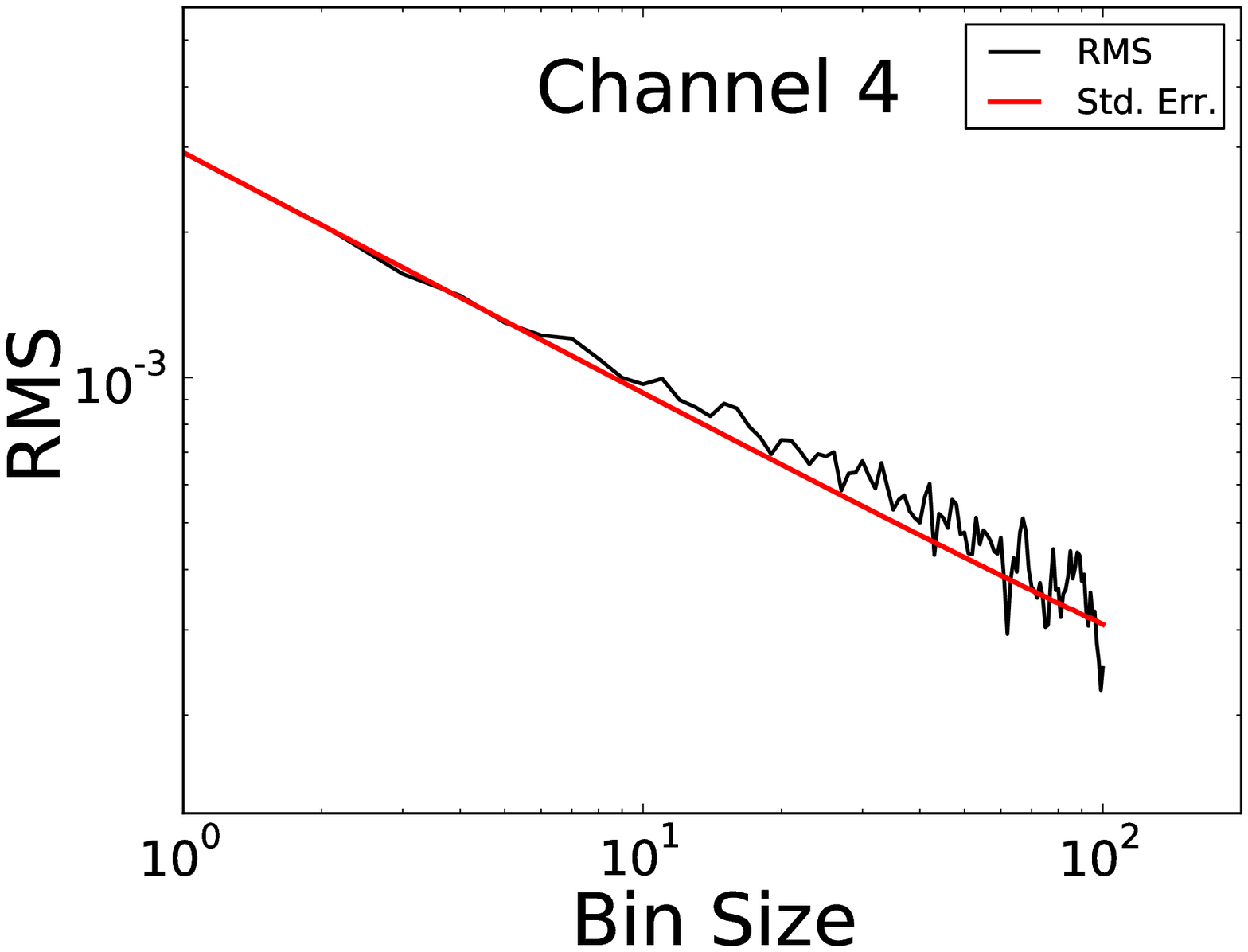}
\figcaption{\label{fig:rms}
Root-mean-squared (RMS) residual flux \textit{vs.}\ bin size in each
channel.  This plot tests for correlated noise.  The straight line is
the prediction for Gaussian white noise.  Since the data do not
deviate far from the line, the effect of correlated noise is minimal.
}
\end{figure}

We also consider the level of correlation in the residuals. For this,
we plot root-mean-squared (RMS) model residuals \textit{vs.} bin size
\citep{PontEtal2006mnrasRedNoise,Winn2008apjTransitLCProj} and compare
to the theoretical \math{1/\sqrt{N}}  RMS scaling.  Figure
\ref{fig:rms} demonstrates the lack of significant photometric noise
correlation in our final models.  In the case of Channel 3, we found a
high degree of correlation between some of the model parameters in the
posterior distribution, and prefer a less-correlated model with
insignificantly poorer BIC and similar SDNR at 3.75 rather than 4.0
pixel aperture size. Differences in interesting parameter values for
the near-optimal alternative are \math{\lesssim 1\sigma}.

\begin{figure}[t]
\if\submitms y
  \setcounter{fignum}{\value{figure}}
  \addtocounter{fignum}{1}
  \newcommand\fignam{f\arabic{fignum}.ps}
\else
  \newcommand\fignam{corr-a.ps}
\fi
\includegraphics[width=\columnwidth, clip]{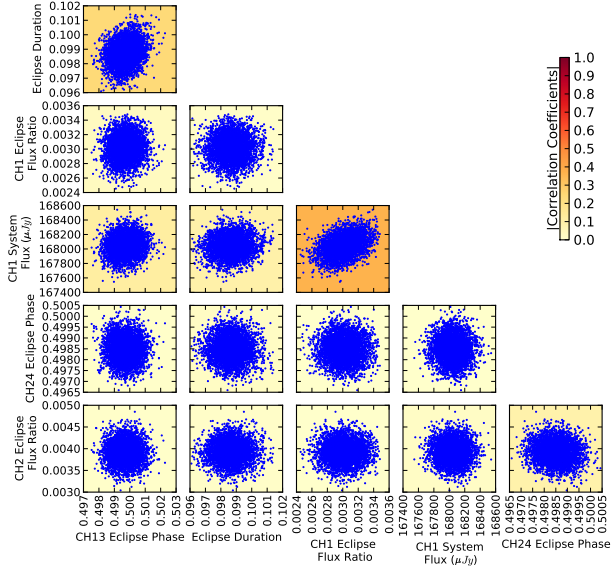}
\figcaption{\label{fig:corr-a}
Parameter correlations.  To decorrelate the Markov chains and
unclutter the plot, one point appears for every 1000th MCMC step.
Each panel contains all the points.
}
\end{figure}

\begin{figure}[t]
\if\submitms y
  \setcounter{fignum}{\value{figure}}
  \addtocounter{fignum}{1}
  \newcommand\fignam{f\arabic{fignum}.ps}
\else
  \newcommand\fignam{corr-b.ps}
\fi
\includegraphics[width=\columnwidth, clip]{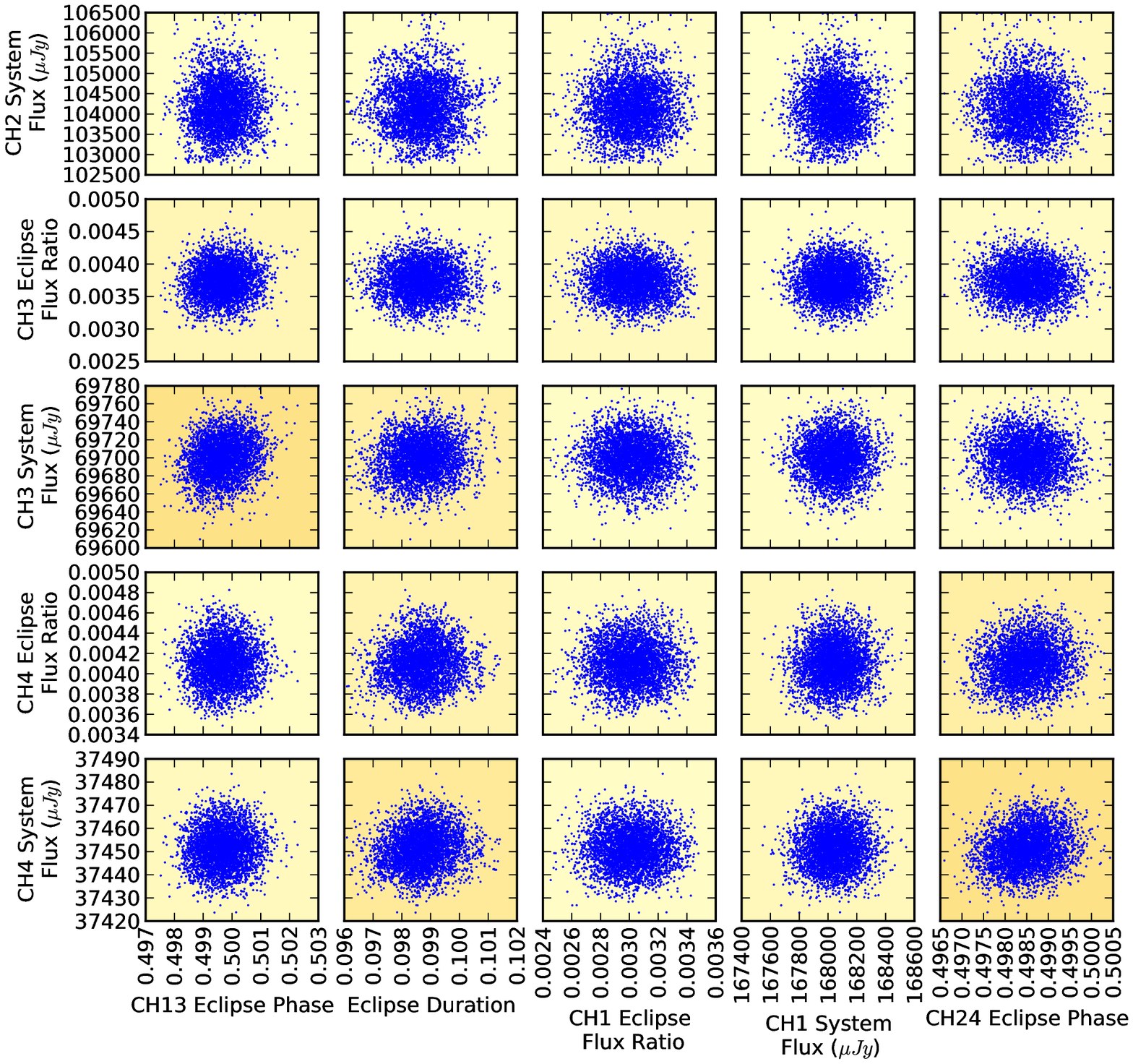}
\figcaption{\label{fig:corr-b}
Parameter correlations, continued.  Same as Figure \ref{fig:corr-a}.
}
\end{figure}

\begin{figure}[t]
\if\submitms y
  \setcounter{fignum}{\value{figure}}
  \addtocounter{fignum}{1}
  \newcommand\fignam{f\arabic{fignum}.ps}
\else
  \newcommand\fignam{corr-c.ps}
\fi
\includegraphics[width=\columnwidth, clip]{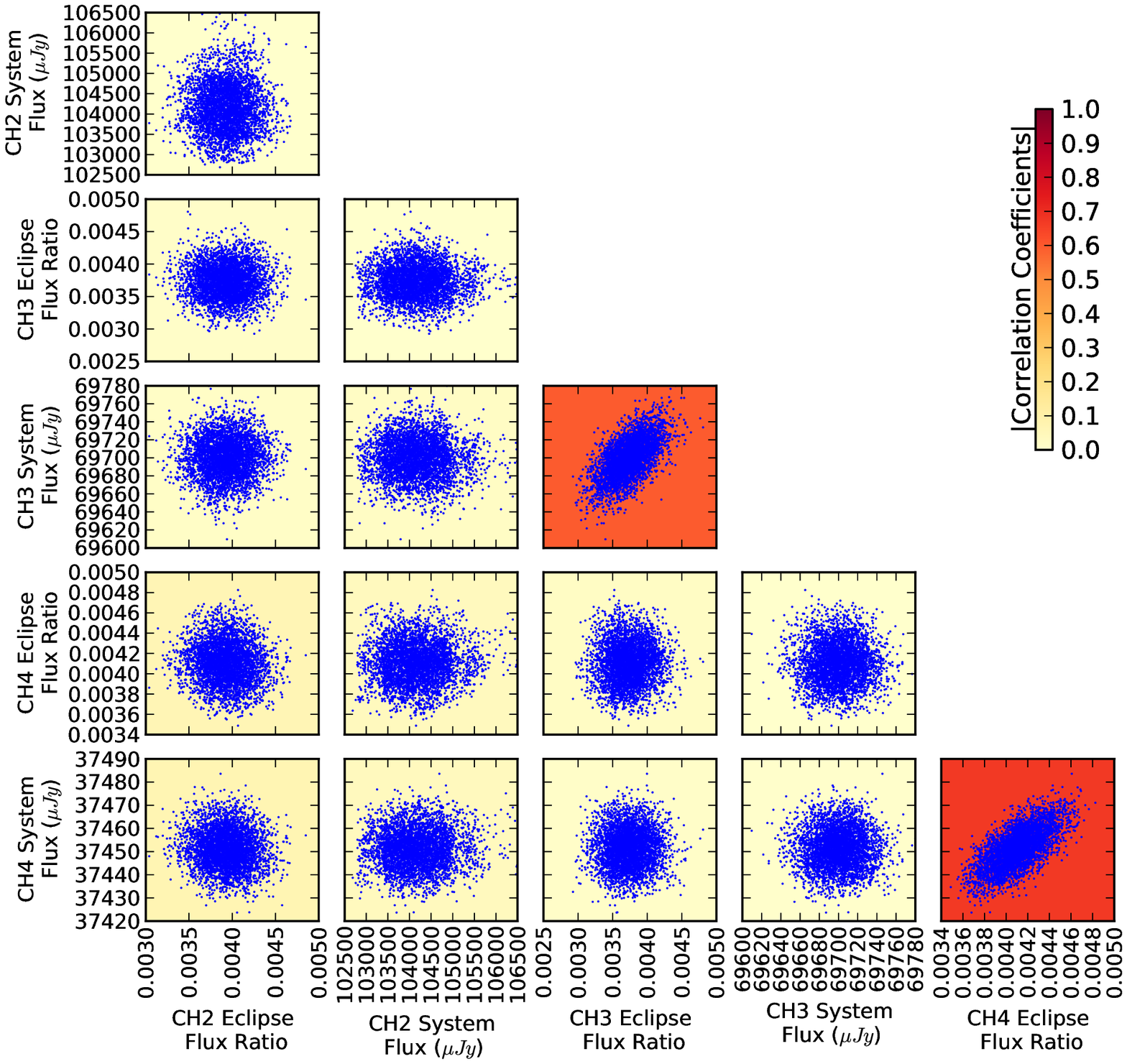}
\figcaption{\label{fig:corr-c}
Parameter correlations, continued.  Same as Figure \ref{fig:corr-a}.
}
\end{figure}

\begin{figure}[t]
\if\submitms y
  \setcounter{fignum}{\value{figure}}
  \addtocounter{fignum}{1}
  \newcommand\fignam{f\arabic{fignum}.ps}
\else
  \newcommand\fignam{hist.ps}
\fi
\includegraphics[width=\columnwidth, clip]{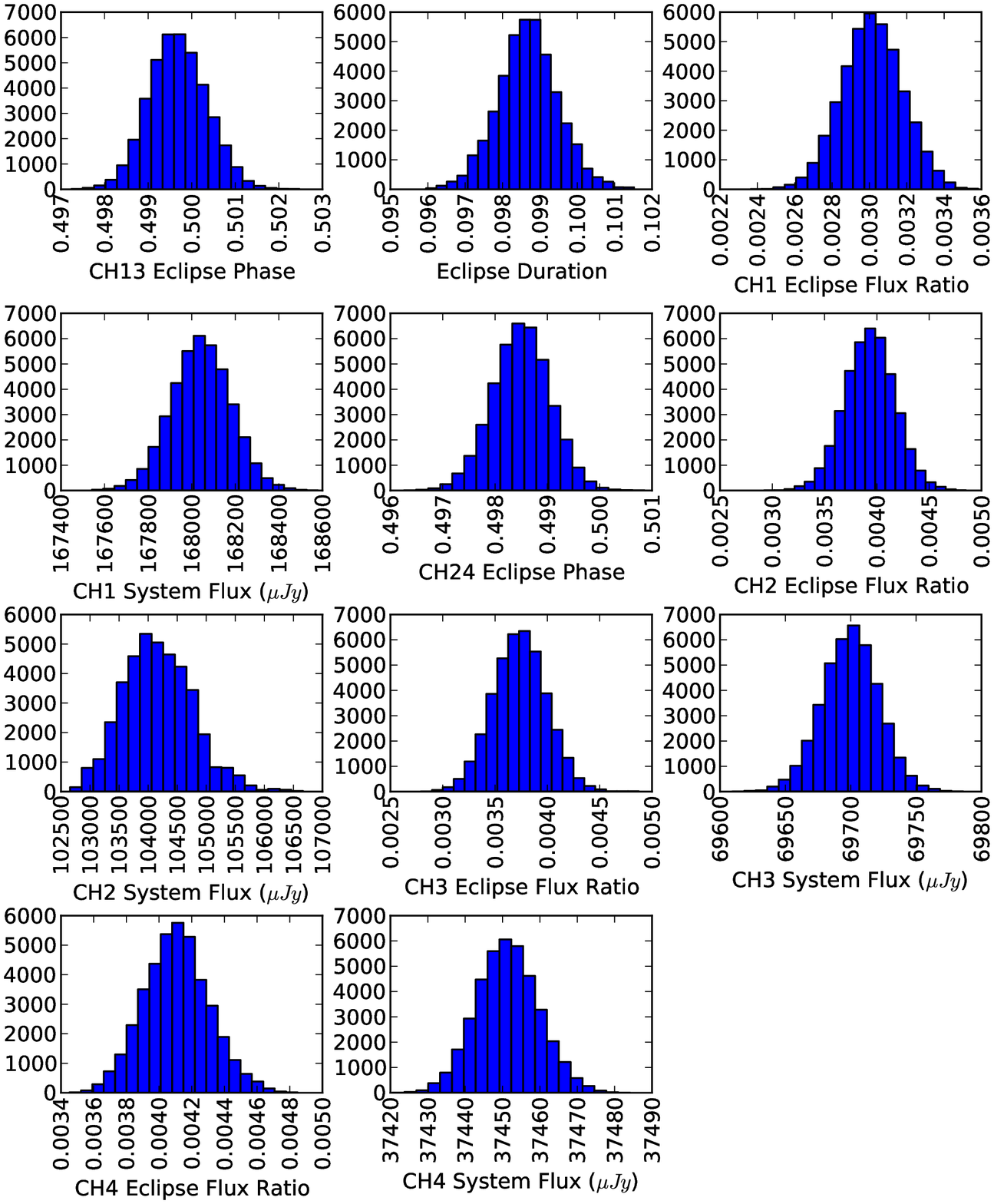}
\figcaption{\label{fig:hist}
Parameter histograms.  To decorrelate the Markov chains, the
histograms come from every 100th MCMC step.
}
\end{figure}

Finally, the marginal posterior distributions (i.e., the parameter
histograms) and plots of their pairwise correlations help in assessing
whether the phase space minimum is global and in determining parameter
uncertainties.  We present these plots for the astrophysical
parameters in Figures \ref{fig:corr-a}, \ref{fig:corr-b},
\ref{fig:corr-c}, and \ref{fig:hist}.  Table \ref{tab:fits} gives the
values and uncertainties of all parameters.

\if\submitms y
\clearpage
\fi
\if\submitms y
\clearpage
\fi

\if\submitms y
\clearpage
\fi
\if\submitms y
\clearpage
\fi

\if\submitms y
\clearpage
\fi
\if\submitms y
\clearpage
\fi

\if\submitms y
\clearpage
\fi
\if\submitms y
\clearpage
\fi

\if\submitms y
\clearpage
\fi
\if\submitms y
\clearpage
\fi

\if\submitms y
\clearpage
\fi
\if\submitms y
\clearpage
\fi

\if\submitms y
\clearpage
\fi
\if\submitms y
\clearpage
\fi

\if\submitms y
\clearpage
\fi
\if\submitms y
\clearpage
\fi

\if\submitms y
\clearpage
\fi
\if\submitms y
\clearpage
\fi

The data files containing the light curves, best-fit models, centering
data, etc., are included as electronic supplements to this
article. Multiple teams analyze the same \textit{Spitzer} exoplanet
data, sometimes obtaining divergent results (e.g.,
\citealp{BeaulieuEtal2010apjGJ436b} and
\citealp{StevensonEtal2010natGJ436b};
\citealp{KnutsonEtal2009apjHD149026bphase} and
\citealp{HarringtonEtal2007natHD149026b}). To facilitate comparison of
these efforts, we encourage all investigators to make similar
disclosure in future reports of exoplanetary transits and
eclipses. Because of differing photometry methods and the vagaries of
estimating error, the standard deviation of the residuals, normalized
to the out-of-eclipse flux, should be the figure of merit for
comparing analyses of the same data by different pipelines. 

\section{ATMOSPHERE}
\label{sec:atm}

We model WASP-18b with the exoplanet atmospheric modeling and
retrieval technique developed by
\citet{MadhusudhanSeager2009apjSpecFit}.  This is a 1D, line-by-line,
radiative-transfer model with constraints of hydrostatic equilibrium
and global energy balance.  The model has six temperature structure
parameters and four molecular abundances, expressed as deviations from
thermochemical equilibrium and solar abundances.  Recognizing the
excess of model parameters over data, our goal is to rule out
unreasonable areas of phase space rather than to determine a unique
composition and thermal profile.  An MCMC routine runs a wide range of
inversion and non-inversion models, integrates the resulting spectra
against the Spitzer bandpasses, and calculates \math{\chi\sp{2}}
  against the four data points.  Integrals over the MCMC posterior
  distribution produce robust statistical statements about the
  underconstrained model.

\begin{figure}[t]
\if\submitms y
  \setcounter{fignum}{\value{figure}}
  \addtocounter{fignum}{1}
  \newcommand\fignam{f\arabic{fignum}.eps}
\else
  \newcommand\fignam{spectra_wasp18.eps}
\fi
\includegraphics[width=\columnwidth, clip]{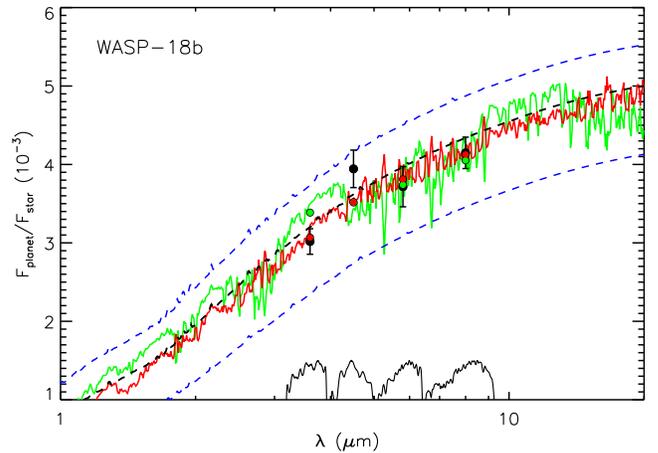}
\figcaption{\label{fig:atm}
Dayside spectrum of WASP-18b.  The black circles with error bars show
our observations of WASP-18b in the four \textit{Spitzer} IRAC
channels.  The red curve shows the inversion model spectrum and the
green curve shows the non-inversion model spectrum discussed in the
text.  The red and green circles are the respective spectra integrated
over the \textit{Spitzer} bandpasses (indicated with arbitrary scale
at the bottom).  The black dashed line shows a blackbody at 3150 K,
and the blue dashed lines show blackbody spectra corresponding to the
minimum and maximum temperatures in the atmosphere (see Figure
\ref{fig:contrib}).
}
\end{figure}

Figure \ref{fig:atm} shows the observed planet-star flux ratios and
two model spectra. We find that the observations can be explained by
models with and without thermal inversions. We note that the
observations are also consistent with a blackbody planetary spectrum
with \math{T =} 3200 K, although a blackbody spectrum is likely
unrealistic.  An atmosphere can have a blackbody spectrum either if it
is isothermal over the upper several optical depths or if there is no
opacity source.  Neither condition is physically favorable.  Several
spectrally active molecules should be abundant in hot-Jupiter
atmospheres and there is collision-induced opacity
\citep{FreedmanEtal2008opacities}.  In addition to being coupled with
the opacities, the temperature structure is also critically influenced
by atmospheric dynamics \citep{ShowmanEtal2009apjAtmospheres}, all of
which can cause a non-isothermal profile.

At 3200 K, the temperature of early M-class stellar photospheres, CO
and H\sb{2}O are the dominant spectroscopically active molecules in
the IR.  Other molecules like CH\sb{4} and CO\sb{2} are negligible,
under the assumption of thermochemical equilibrium with solar
abundances.  CO has a strong absorption feature in the 4.5 {\micron}
channel.  H\sb{2}O contributes the dominant opacity in the 5.8
{\micron} channel, and contributes significantly in the remaining
channels.  Thus, for temperature decreasing monotonically with
altitude, i.e., in the absence of a thermal inversion, the spectra
should exhibit noticeable absorption in the 4.5 and 5.8 {\micron}
channels and less in the 3.6 {\micron} and 8 {\micron} channels
\citep{MadhusudhanSeager2010apjThermInv,
  MadhusudhanSeager2011apjGJ436b}.  The observed planet-star flux
contrast in the 4.5 {\micron} channel should then be lower than that
in the 3.6 {\micron} channel
\citep{CharbonneauEtal2008apjHD189733bbroad,
  StevensonEtal2010natGJ436b}; the difference depends on the
temperature gradient and the composition.

\begin{figure}[t]
\if\submitms y
  \setcounter{fignum}{\value{figure}}
  \addtocounter{fignum}{1}
  \newcommand\fignam{f\arabic{fignum}.eps}
\else
  \newcommand\fignam{contrib.eps}
\fi
\includegraphics[width=\columnwidth, clip]{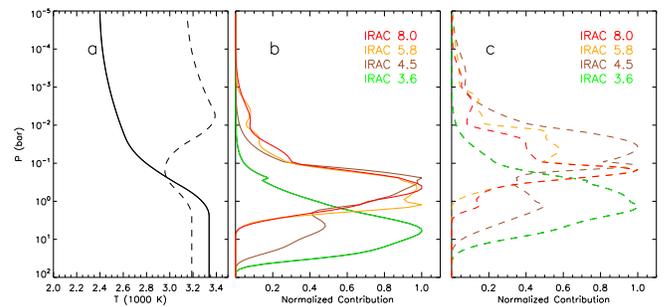}
\figcaption{\label{fig:contrib}
Pressure-temperature profile (left) and contribution functions (middle
and right). The middle and right panels show the normalized
contribution functions for the non-inversion and inversion models,
respectively, in the indicated \textit{Spitzer} filters, with
wavelengths in {\micron}.  The left panel overlays the profiles for
both models.
}
\end{figure}

Our observations of WASP-18b show excess flux at 4.5 {\micron},
compared to the 3.6 {\micron} channel.  This could be due to a thermal
inversion, but the observational uncertainties also allow just a
gentle temperature gradient and no inversion, such that the absorption
features are not too deep, and a different chemical composition
\citep{MadhusudhanSeager2010apjThermInv}.  Two models, with and
without an inversion, appear in Figure \ref{fig:atm}.  Both explain
the data fairly well, the inversion model at the 1\math{\sigma} level
and the non-inversion model within 1.5\math{\sigma}.  The molecular
abundances of the models are only marginally different from those of
thermochemical equilibrium with solar abundances
(TE\sb{\math{\odot}}).  The inversion model has 10 times more CO, and
the non-inversion model has 10 times less H\sb{2}O and CO, as compared
to TE\sb{\math{\odot}}.  The mixing ratio of CO\sb{2} is \ttt{-7} for
the inversion model and \ttt{-8} for the non-inversion model.  Despite
the weak constraints on the temperature structure, the observations do
place a strict constraint on the day-night energy redistribution in
WASP-18b: Both models require a low Bond albedo (\math{A}) and
inefficient day-night redistribution (\math{\lesssim 0.1} for
\math{A}=0) in WASP-18b.  Figure \ref{fig:contrib} shows the
contribution functions of the two models in the four IRAC channels,
along with the thermal profiles.

The presence of a thermal inversion in the dayside atmosphere of
WASP-18b is expected based on theoretical grounds
\citep{HubenyEtal2003apjbifurcation, FortneyEtal2008TiOVO}.  At the
high temperatures of this planet, TiO and VO can exist in gas phase
over the entire atmosphere, thus contributing to the strong visible
opacities required to form stratospheres.  However, questions of
whether the concentrations of TiO/VO alone are adequate to cause the
required thermal inversion, and whether other sources of visible/UV
opacity are possible at these temperatures, merit future theoretical
investigation \citep{SpiegelEtal2009apjTiO, ZahnleEtal2009Soot}.
Furthermore, \citet{KnutsonEtal2010apjStellarActivity} find that the
host star WASP-18 has a low activity level, indicating that
inversion-causing compounds are not likely to be destroyed by stellar
UV radiation, thereby also favoring the presence of a thermal
inversion.  Thus, WASP-18b is an apt candidate for follow-up
observations searching for thermal inversions.  Stronger constraints
on the temperature structure of WASP-18b are possible in the near
future if ground-based observations of thermal emission become
available \citep{MadhusudhanEtal2011NatureWASP12b}.  Also, the low
day-night energy redistribution required by the present observations
can be verified by potential future observations of thermal phase
curves (e.g., \citep{KnutsonEtal2008apjHD209458b}) of WASP-18b.

\if\submitms y
\clearpage
\fi
\if\submitms y
\clearpage
\fi

\if\submitms y
\clearpage
\fi
\if\submitms y
\clearpage
\fi

\section{ORBITAL DYNAMICS}
\label{sec:dyn}

Our secondary eclipse times further constrain the planet's
already-precise orbital parameters.
\citet{Triaudetal2010apjSpinOrbitMeas} detect an eccentricity for
WASP-18b of \math{e = 0.0085 \pm{0.0008}}, the lowest fully-determined
value for any transiting planet measured with such precision.  A joint
photometric fit to all four \textit{Spitzer} observations yields a
midpoint phase of \math{0.4990 \pm{0.0004}} for the
\citet{HellierEtal2009natWASP18b} ephemeris, and a duration of
\math{D\sb{\rm s} = 0.0927\pm{0.0007}} days, which is longer than the
transit duration of \citet{HellierEtal2009natWASP18b} by 4.8
\math{\sigma}.  By itself, after a 20-second light-time correction,
the eclipse midpoint tells us that \math{e \cos \omega = -0.0016
  \pm{0.0007}}, where \math{\omega} is the longitude of periastron.
We combine the eclipse phase and duration with known transit
parameters from \citet{HellierEtal2009natWASP18b} to determine that
\math{e \sin \omega = 0.0198 \pm 0.0072}.  To do this, we use Eq.\ 18
from \citet{RagozzineEtal2009apjTransitLightCurves} as derived by
\citet{Koppal1959BinarySystems}:
 
\begin{eqnarray}
e \sin \omega =
 \left(\frac{D\sb{\rm S}-D\sb{\rm P}}{D\sb{\rm S}+D\sb{\rm P}}\right)
 \left(\frac{\alpha{\sp 2}-\cos{\sp 2}i}{\alpha{\sp 2}-2\cos\sp{2}i}\right).
\end{eqnarray}

\noindent \math{D\sb{\rm P}} is the transit duration and
\math{D\sb{\rm S}} is the secondary eclipse duration. We define
\math{\alpha = \frac{R\sb{\rm *}}{a}(1+\frac{R\sb{\rm
      p}}{R\sb{\rm*}})\frac{1}{\sqrt{1-e\sp{2}}}} and \math{\cos i =
  b\sb{\rm P}\frac{R\sb{\rm *}}{a}\frac{1+e \sin \omega}{1-e\sp{2}}},
and \math{\frac{R\sb{*}}{a} = \frac{D\sb{{\rm
        p}}}{P}\frac{\pi}{\sqrt{(\frac{R\sb{\rm p}}{R\sb{\rm
          *}}+1)\sp{2}-b\sb{\rm p}\sp{2}}}}, where \math{R\sb{\rm p}}
and \math{R\sb{\rm *}} are the planetary and stellar radii,
respectively, \math{a} is the orbit's semi-major axis, \math{P} is the
orbital period, and \math{b\sb{\rm P}} is the impact parameter in
units of the stellar radius.  We solved the equation numerically for
\math{e \sin \omega}, and the uncertainties come from sampling
Gaussian distributions generated from the uncertainties of the input
parameters.

We jointly fit an MCMC orbit model \citep{CampoEtal2010apjWASP18b} to
the BJD time for 2 eclipses, 37 radial velocity (RV) data points from
\citet{Triaudetal2010apjSpinOrbitMeas}, and 6 transit midpoints
extracted from the photometry of \citet{SouthworthEtal2009apjWASP18b},
omitting three RV points subject to the Rossiter-McLaughlin
effect.  This results in a total of 45 data points, 42 of which are
included.  Our model has 6 free parameters. Results are given in Table \ref{tab:orbit}, where
\math{T\sb{0}} is the ephemeris time, \math{K} is the RV amplitude,
and \math{\gamma} is the barycentric velocity.

We find from the fit above that \math{e \cos \omega = -0.00014 {\pm}
  0.00053}, consistent with 0, leaving \math{e} dominated by its
\math{e \sin \omega} component.  We note that the value of \math{e
  \sin \omega} found photometrically is positive, while the value in
our joint fit is negative.  Eclipse timing does not effectively
constrain \math{ e \sin \omega}.  Radial velocity measurements have a
known tendency to overestimate \math{e} when it is low
\citep{LaughlinEtal2005apjHD209458orbit}.  A key sign of this is a
value of \math{\omega \sim \pm 90 \degrees}.  Photometric information,
such as the measured durations for our eclipses, cannot yet
independently confirm a non-zero \math{ e \sin \omega} beyond the
3\math{\sigma} level.  Precise determination of \math{e} is important
because the circularization timescale for a tidal damping quality
factor, \math{Q\sb{\rm p}} {\sim} 10\sp{6}
\citep{Mardling2007mnrasTidalEvo}, is comparable to the age of the
system.  Whether the orbit is still eccentric determines if it is
still experiencing tidal dissipation, which drives the evolution of
the system.  We perform a comparison fit with \math{e = 0}, but its
BIC value of 155 is considerably higher than BIC = 102 for the
eccentric fit.

To determine if the eccentricity we found could have come up by random
chance, we also performed an experiment similar to that of
\citet{LaughlinEtal2005apjHD209458orbit} in which we generated
\math{10\sp{5}} radial velocity datasets for a planet in a circular
orbit with the same period, mid-transit time, and semi-amplitude as
our best fit for WASP-18b. In each dataset, the BJD of each
observation was kept the same as in the real WASP-18b dataset. We
added Gaussian noise corresponding to the instrumental error for each
observation and a 3 m s\sp{-1} stellar jitter consistent with the
tables of \citet{Wright2005paspJitter}, in quadrature.  We retained
the transit and eclipse timings.  We then used a minimizer to find
combinations of \math{e \sin \omega} and \math{e \cos \omega} that
minimize the \math{\chi\sp{2}} corresponding to each dataset.  The
mean and standard deviation of \math{e \sin \omega} and \math{e \cos
  \omega} were \math{0.0000 \pm 0.0012} and \math{-0.0003 \pm 0.0002},
respectively. The \math{3\sigma} upper limit on \math{|e \sin \omega|}
is \math{0.0036}, well below our best-fit value in the real
dataset. This is not surprising because of the high signal-to-noise
ratio and even sampling \citep{Triaudetal2010apjSpinOrbitMeas} of the
RV data (\math{\frac{K}{\langle{\sigma}\sb{RV}\rangle} \sim
  180}). Given the MCMC results, the BIC comparison between the fits
to a circular and non-circular orbit, the Monte Carlo experiments, and
the weak but consistent photometric support, and despite the
improbable orientation of the orbit, the 7\math{\sigma} non-zero
eccentricity of WASP-18b's orbit is likely not an overestimate.

Despite WASP-18b's short period and close proximity to its host star,
its high density and low eccentricity make it a poor candidate for the
detection of apsidal precession
\citep{RagozzineEtal2009apjTransitLightCurves,CampoEtal2010apjWASP18b}.
The precession should manifest itself as an eclipse/transit timing
variation with a period of 600 years and an amplitude of
\math{\frac{eP}{\pi}}, or about four minutes
\citep{RagozzineEtal2009apjTransitLightCurves}.  Given the orbit's
current orientation, apsidal motion could be detectable as a \sim 7 ms
difference between the best-fit transit and eclipse periods
\citep{GimenezEtal1995apsEclipsingBinaries}.  This signal would likely
be overwhelmed by the modulation of the period due to tidal infall
\citep{HellierEtal2009natWASP18b}, which could be measurable within a
few decades. The Applegate effect makes very small transit timing
variations unmeasurable \citep{WatsonEtal2010arxiv}.  

\if\submitms y
\clearpage
\fi
\atabon\begin{deluxetable}{lr@{\,{\pm}\,}lr@{\,\,}}
\tablecaption{\label{tab:orbit} Joint Orbital Fits}
\tablewidth{0pt}

\tablehead{
\colhead{Parameter} &
\mctc{Value} &
}

\startdata
\math{e \sin \omega}              	&       0.0091    & 0.0012    \\
\math{e \cos \omega}              	& \math{-0.00014} & 0.00053   \\
\math{e}                                &     0.0091      & 0.0012    \\
\math{\omega} (\degree)           	& \math{-91}      & 3         \\
\math{P} (days)                   	&     0.9414518   & 0.0000004 \\
\math{T\sb{0}} (MJD)\tablenotemark{a}   &  1084.79363     & 0.00011   \\
\math{K} (ms\sp{-1})                    &  1818           & 3         \\
\math{\gamma} (ms\sp{-1})               &  3327           & 2         \\
BIC                                     &  \mctc{102.0}              \         
\enddata

\tablenotetext{a}{MJD = BJD - 2,454,000 (Terrestrial Time)}
\end{deluxetable}\ataboff
\if\submitms y
\clearpage
\fi
\placetable{tab:orbit}

\section{CONCLUSIONS}
\label{sec:concl}

\textit{Spitzer} observed two secondary eclipses of WASP-18b using all
four channels of the IRAC instrument. A blackbody model fits the
observed brightness temperatures relatively well.  Slightly better
fits to both inversion and non-inversion models exist with the
inversion model somewhat preferred.  Because the planet is so much
brighter than its predicted equilibrium temperature for uniform
redistribution, the model requires near-zero albedo and very low
day-night energy redistribution. The very small scale height makes
this atmosphere interesting as an extreme example among irradiated
planets. The addition of secondary eclipse data also improves the
orbital parameters, confirming a slight eccentricity.  Files
containing the lightcurves, model fits, source centers, and other
ancillary data appear as electronic attachments to this article.
  
\acknowledgments We thank Drake Deming for helpful discussions.  We received free
software and services from SciPy, Matplotlib, and the Python
Programming Language community; W.\ Landsman and other contributors to
the Interactive Data Language Astronomy Library, the free and
open-source software communities; the NASA Astrophysics Data System,
and the JPL Solar System Dynamics group.  This work is based in part
on observations made with the \textit{Spitzer Space Telescope}, which
is operated by the Jet Propulsion Laboratory, California Institute of
Technology under a contract with NASA.  Support for this work was
provided by NASA through an award issued by JPL/Caltech.

\appendix

\nojoe{
\if\submitms y
  \newcommand\bblnam{ms}
\else
  \newcommand\bblnam{wasp18}
\fi
\bibliography{\bblnam}}

\begin{thebibliography}{46}
\expandafter\ifx\csname natexlab\endcsname\relax\def\natexlab#1{#1}\fi

\bibitem[{{Beaulieu} {et~al.}(2011){Beaulieu}, {Tinetti}, {Kipping}, {Ribas},
  {Barber}, {Cho}, {Polichtchouk}, {Tennyson}, {Yurchenko}, {Griffith},
  {Batista}, {Waldmann}, {Miller}, {Carey}, {Mousis}, {Fossey}, \&
  {Aylward}}]{BeaulieuEtal2010apjGJ436b}
{Beaulieu}, J.-P. {et~al.} 2011, \apj, 731, 16, arXiv:1007.0324

\bibitem[{{Burrows} {et~al.}(2008){Burrows}, {Ibgui}, \&
  {Hubeny}}]{BurrowsEtal2008apjHD209458b}
{Burrows}, A., {Ibgui}, L., \& {Hubeny}, I. 2008, \apj, 682, 1277,
  arXiv:0803.2523

\bibitem[{{Campo} {et~al.}(2011){Campo}, {Harrington}, {Hardy}, {Stevenson},
  {Nymeyer}, {Ragozzine}, {Lust}, {Anderson}, {Collier-Cameron}, {Blecic},
  {Britt}, {Bowman}, {Wheatley}, {Loredo}, {Deming}, {Hebb}, {Hellier},
  {Maxted}, {Pollaco}, \& {West}}]{CampoEtal2010apjWASP18b}
{Campo}, C.~J. {et~al.} 2011, \apj, 727, 125, arXiv:1003.2763

\bibitem[{{Charbonneau} {et~al.}(2005){Charbonneau}, {Allen}, {Megeath},
  {Torres}, {Alonso}, {Brown}, {Gilliland}, {Latham}, {Mandushev}, {O'Donovan},
  \& {Sozzetti}}]{CharbonneauEtal2005apjTrES1}
{Charbonneau}, D. {et~al.} 2005, \apj, 626, 523, arXiv:astro-ph/0503457

\bibitem[{{Charbonneau} {et~al.}(2008){Charbonneau}, {Knutson}, {Barman},
  {Allen}, {Mayor}, {Megeath}, {Queloz}, \&
  {Udry}}]{CharbonneauEtal2008apjHD189733bbroad}
{Charbonneau}, D., {Knutson}, H.~A., {Barman}, T., {Allen}, L.~E., {Mayor}, M.,
  {Megeath}, S.~T., {Queloz}, D., \& {Udry}, S. 2008, \apj, 686, 1341,
  arXiv:0802.0845

\bibitem[{{Deming} {et~al.}(2005){Deming}, {Seager}, {Richardson}, \&
  {Harrington}}]{DemingEtal2005natHD209458b}
{Deming}, D., {Seager}, S., {Richardson}, L.~J., \& {Harrington}, J. 2005,
  \nat, 434, 740, arXiv:astro-ph/0503554

\bibitem[{{Fazio} {et~al.}(2004){Fazio}, {Hora}, {Allen}, {Ashby}, {Barmby},
  {Deutsch}, {Huang}, {Kleiner}, {Marengo}, {Megeath}, {Melnick}, {Pahre},
  {Patten}, {Polizotti}, {Smith}, {Taylor}, {Wang}, {Willner}, {Hoffmann},
  {Pipher}, {Forrest}, {McMurty}, {McCreight}, {McKelvey}, {McMurray}, {Koch},
  {Moseley}, {Arendt}, {Mentzell}, {Marx}, {Losch}, {Mayman}, {Eichhorn},
  {Krebs}, {Jhabvala}, {Gezari}, {Fixsen}, {Flores}, {Shakoorzadeh}, {Jungo},
  {Hakun}, {Workman}, {Karpati}, {Kichak}, {Whitley}, {Mann}, {Tollestrup},
  {Eisenhardt}, {Stern}, {Gorjian}, {Bhattacharya}, {Carey}, {Nelson},
  {Glaccum}, {Lacy}, {Lowrance}, {Laine}, {Reach}, {Stauffer}, {Surace},
  {Wilson}, {Wright}, {Hoffman}, {Domingo}, \& {Cohen}}]{FazioEtal2004apjsIRAC}
{Fazio}, G.~G. {et~al.} 2004, \apjs, 154, 10, arXiv:astro-ph/0405616

\bibitem[{{Fortney} {et~al.}(2008){Fortney}, {Lodders}, {Marley}, \&
  {Freedman}}]{FortneyEtal2008TiOVO}
{Fortney}, J.~J., {Lodders}, K., {Marley}, M.~S., \& {Freedman}, R.~S. 2008,
  \apj, 678, 1419, arXiv:0710.2558

\bibitem[{{Freedman} {et~al.}(2008){Freedman}, {Marley}, \&
  {Lodders}}]{FreedmanEtal2008opacities}
{Freedman}, R.~S., {Marley}, M.~S., \& {Lodders}, K. 2008, \apjs, 174, 504,
  arXiv:0706.2374

\bibitem[{{Gim{\'e}nez} \&
  {Bastero}(1995)}]{GimenezEtal1995apsEclipsingBinaries}
{Gim{\'e}nez}, A., \& {Bastero}, M. 1995, \apss, 226, 99

\bibitem[{{Harrington} {et~al.}(2007){Harrington}, {Luszcz}, {Seager},
  {Deming}, \& {Richardson}}]{HarringtonEtal2007natHD149026b}
{Harrington}, J., {Luszcz}, S.~H., {Seager}, S., {Deming}, D., \& {Richardson},
  L.~J. 2007, \nat, 447, 691

\bibitem[{{Hellier} {et~al.}(2009){Hellier}, {Anderson}, {Cameron}, {Gillon},
  {Hebb}, {Maxted}, {Queloz}, {Smalley}, {Triaud}, {West}, {Wilson}, {Bentley},
  {Enoch}, {Horne}, {Irwin}, {Lister}, {Mayor}, {Parley}, {Pepe}, {Pollacco},
  {Segransan}, {Udry}, \& {Wheatley}}]{HellierEtal2009natWASP18b}
{Hellier}, C. {et~al.} 2009, \nat, 460, 1098

\bibitem[{{Hubeny} {et~al.}(2003){Hubeny}, {Burrows}, \&
  {Sudarsky}}]{HubenyEtal2003apjbifurcation}
{Hubeny}, I., {Burrows}, A., \& {Sudarsky}, D. 2003, \apj, 594, 1011,
  arXiv:astro-ph/0305349

\bibitem[{{Knutson} {et~al.}(2008){Knutson}, {Charbonneau}, {Allen}, {Burrows},
  \& {Megeath}}]{KnutsonEtal2008apjHD209458b}
{Knutson}, H.~A., {Charbonneau}, D., {Allen}, L.~E., {Burrows}, A., \&
  {Megeath}, S.~T. 2008, \apj, 673, 526, arXiv:0709.3984

\bibitem[{{Knutson} {et~al.}(2009{\natexlab{a}}){Knutson}, {Charbonneau},
  {Burrows}, {O'Donovan}, \& {Mandushev}}]{KnutsonEtal2009apjTres4}
{Knutson}, H.~A., {Charbonneau}, D., {Burrows}, A., {O'Donovan}, F.~T., \&
  {Mandushev}, G. 2009{\natexlab{a}}, \apj, 691, 866, arXiv:0810.0021

\bibitem[{{Knutson} {et~al.}(2009{\natexlab{b}}){Knutson}, {Charbonneau},
  {Cowan}, {Fortney}, {Showman}, {Agol}, \&
  {Henry}}]{KnutsonEtal2009apjHD149026bphase}
{Knutson}, H.~A., {Charbonneau}, D., {Cowan}, N.~B., {Fortney}, J.~J.,
  {Showman}, A.~P., {Agol}, E., \& {Henry}, G.~W. 2009{\natexlab{b}}, \apj,
  703, 769, arXiv:0908.1977

\bibitem[{{Knutson} {et~al.}(2010){Knutson}, {Howard}, \&
  {Isaacson}}]{KnutsonEtal2010apjStellarActivity}
{Knutson}, H.~A., {Howard}, A.~W., \& {Isaacson}, H. 2010, \apj, 720, 1569,
  arXiv:1004.2702

\bibitem[{{Kopal}(1959)}]{Koppal1959BinarySystems}
{Kopal}, Z. 1959, Close binary systems (The International Astrophysics Series;
  London: Chapman \& Hall)

\bibitem[{{Laughlin} {et~al.}(2005){Laughlin}, {Marcy}, {Vogt}, {Fischer}, \&
  {Butler}}]{LaughlinEtal2005apjHD209458orbit}
{Laughlin}, G., {Marcy}, G.~W., {Vogt}, S.~S., {Fischer}, D.~A., \& {Butler},
  R.~P. 2005, \apjl, 629, L121

\bibitem[{{Liddle}(2007)}]{Liddle2007apjInfoCrit}
{Liddle}, A.~R. 2007, \mnras, 377, L74, arXiv:astro-ph/0701113

\bibitem[{{Machalek} {et~al.}(2008){Machalek}, {McCullough}, {Burke},
  {Valenti}, {Burrows}, \& {Hora}}]{MachalekEtal2008XO-1b}
{Machalek}, P., {McCullough}, P.~R., {Burke}, C.~J., {Valenti}, J.~A.,
  {Burrows}, A., \& {Hora}, J.~L. 2008, \apj, 684, 1427, arXiv:0805.2418

\bibitem[{{Machalek} {et~al.}(2009){Machalek}, {McCullough}, {Burrows},
  {Burke}, {Hora}, \& {Johns-Krull}}]{MachalekEtal2009XO-2b}
{Machalek}, P., {McCullough}, P.~R., {Burrows}, A., {Burke}, C.~J., {Hora},
  J.~L., \& {Johns-Krull}, C.~M. 2009, \apj, 701, 514, arXiv:0906.1293

\bibitem[{{Machalek} {et~al.}(2011)}]{MachalekEtal2010}
{Machalek}, P., {et~al.} 2011, \apj, submitted

\bibitem[{{Madhusudhan} {et~al.}(2011){Madhusudhan}, {Harrington}, {Stevenson},
  {Nymeyer}, {Campo}, {Wheatley}, {Deming}, {Blecic}, {Hardy}, {Lust},
  {Anderson}, {Collier-Cameron}, {Britt}, {Bowman}, {Hebb}, {Hellier},
  {Maxted}, {Pollacco}, \& {West}}]{MadhusudhanEtal2011NatureWASP12b}
{Madhusudhan}, N. {et~al.} 2011, \nat, 469, 64, arXiv:1012.1603

\bibitem[{{Madhusudhan} \& {Seager}(2009)}]{MadhusudhanSeager2009apjSpecFit}
{Madhusudhan}, N., \& {Seager}, S. 2009, \apj, 707, 24, arXiv:0910.1347

\bibitem[{{Madhusudhan} \& {Seager}(2010)}]{MadhusudhanSeager2010apjThermInv}
------. 2010, \apj, 725, 261, arXiv:1010.4585

\bibitem[{{Madhusudhan} \& {Seager}(2011)}]{MadhusudhanSeager2011apjGJ436b}
------. 2011, \apj, 729, 41, arXiv:1004.5121

\bibitem[{{Mandel} \& {Agol}(2002)}]{MandelAgol2002apjLightcurves}
{Mandel}, K., \& {Agol}, E. 2002, \apjl, 580, L171, arXiv:astro-ph/0210099

\bibitem[{{Mardling}(2007)}]{Mardling2007mnrasTidalEvo}
{Mardling}, R.~A. 2007, \mnras, 382, 1768, arXiv:0706.0224

\bibitem[{{O'Donovan} {et~al.}(2010){O'Donovan}, {Charbonneau}, {Harrington},
  {Madhusudhan}, {Seager}, {Deming}, \& {Knutson}}]{ODonovanEtal2010apjTres-2b}
{O'Donovan}, F.~T., {Charbonneau}, D., {Harrington}, J., {Madhusudhan}, N.,
  {Seager}, S., {Deming}, D., \& {Knutson}, H.~A. 2010, \apj, 710, 1551,
  arXiv:0909.3073

\bibitem[{{Pont} {et~al.}(2006){Pont}, {Zucker}, \&
  {Queloz}}]{PontEtal2006mnrasRedNoise}
{Pont}, F., {Zucker}, S., \& {Queloz}, D. 2006, \mnras, 373, 231,
  arXiv:astro-ph/0608597

\bibitem[{{Ragozzine} \& {Wolf}(2009)}]{RagozzineEtal2009apjTransitLightCurves}
{Ragozzine}, D., \& {Wolf}, A.~S. 2009, \apj, 698, 1778, arXiv:0807.2856

\bibitem[{{Seager} \&
  {Mall{\'e}n-Ornelas}(2003)}]{SeagerEtal2003apjPlanetStarParameters}
{Seager}, S., \& {Mall{\'e}n-Ornelas}, G. 2003, \apj, 585, 1038,
  arXiv:astro-ph/0206228

\bibitem[{{Showman} {et~al.}(2009){Showman}, {Fortney}, {Lian}, {Marley},
  {Freedman}, {Knutson}, \& {Charbonneau}}]{ShowmanEtal2009apjAtmospheres}
{Showman}, A.~P., {Fortney}, J.~J., {Lian}, Y., {Marley}, M.~S., {Freedman},
  R.~S., {Knutson}, H.~A., \& {Charbonneau}, D. 2009, \apj, 699, 564,
  arXiv:0809.2089

\bibitem[{{Sing} \& {L{\'o}pez-Morales}(2009)}]{SingEtal2009aaGroundbased}
{Sing}, D.~K., \& {L{\'o}pez-Morales}, M. 2009, \aap, 493, L31, arXiv:0901.1876

\bibitem[{{Southworth} {et~al.}(2009){Southworth}, {Hinse}, {Dominik},
  {Glitrup}, {J{\o}rgensen}, {Liebig}, {Mathiasen}, {Anderson}, {Bozza},
  {Browne}, {Burgdorf}, {Calchi Novati}, {Dreizler}, {Finet}, {Harps{\o}e},
  {Hessman}, {Hundertmark}, {Maier}, {Mancini}, {Maxted}, {Rahvar}, {Ricci},
  {Scarpetta}, {Skottfelt}, {Snodgrass}, {Surdej}, \&
  {Zimmer}}]{SouthworthEtal2009apjWASP18b}
{Southworth}, J. {et~al.} 2009, \apj, 707, 167, arXiv:0910.4875

\bibitem[{{Southworth} {et~al.}(2007){Southworth}, {Wheatley}, \&
  {Sams}}]{SouthworthEtal2007mnrasHD209458b}
{Southworth}, J., {Wheatley}, P.~J., \& {Sams}, G. 2007, \mnras, 379, L11,
  arXiv:0704.1570

\bibitem[{{Spiegel} {et~al.}(2009){Spiegel}, {Silverio}, \&
  {Burrows}}]{SpiegelEtal2009apjTiO}
{Spiegel}, D.~S., {Silverio}, K., \& {Burrows}, A. 2009, \apj, 699, 1487,
  arXiv:0902.3995

\bibitem[{{Stevenson} {et~al.}(2010){Stevenson}, {Harrington}, {Nymeyer},
  {Madhusudhan}, {Seager}, {Bowman}, {Hardy}, {Deming}, {Rauscher}, \&
  {Lust}}]{StevensonEtal2010natGJ436b}
{Stevenson}, K.~B. {et~al.} 2010, \nat, 464, 1161, arXiv:1010.4591v1

\bibitem[{{Triaud} {et~al.}(2010){Triaud}, {Collier Cameron}, {Queloz},
  {Anderson}, {Gillon}, {Hebb}, {Hellier}, {Loeillet}, {Maxted}, {Mayor},
  {Pepe}, {Pollacco}, {S{\'e}gransan}, {Smalley}, {Udry}, {West}, \&
  {Wheatley}}]{Triaudetal2010apjSpinOrbitMeas}
{Triaud}, A.~H.~M.~J. {et~al.} 2010, \aap, 524, A25, arXiv:1008.2353

\bibitem[{{Watson} \& {Marsh}(2010)}]{WatsonEtal2010arxiv}
{Watson}, C.~A., \& {Marsh}, T.~R. 2010, \mnras, 601, arXiv:1003.0340

\bibitem[{{Wheatley} {et~al.}(2011){Wheatley}, {Collier Cameron}, {Harrington},
  {Fortney}, {Simpson}, {Anderson}, {Smith}, {Aigrain}, {Clarkson}, {Gillon},
  {Haswell}, {Hebb}, {H{\'e}brard}, {Hellier}, {Hodgkin}, {Horne}, {Kane},
  {Maxted}, {Norton}, {Pollacco}, {Pont}, {Skillen}, {Smalley}, {Street},
  {Udry}, {West}, \& {Wilson}}]{WheatleyEtal2010}
{Wheatley}, P.~J. {et~al.} 2011, \apj, submitted, arXiv:1004.0836

\bibitem[{{Winn} {et~al.}(2008){Winn}, {Holman}, {Torres}, {McCullough},
  {Johns-Krull}, {Latham}, {Shporer}, {Mazeh}, {Garcia-Melendo}, {Foote},
  {Esquerdo}, \& {Everett}}]{Winn2008apjTransitLCProj}
{Winn}, J.~N. {et~al.} 2008, \apj, 683, 1076, arXiv:0804.4475

\bibitem[{{Wright}(2005)}]{Wright2005paspJitter}
{Wright}, J.~T. 2005, \pasp, 117, 657, arXiv:astro-ph/0505214

\bibitem[{{Zahnle} {et~al.}(2009{\natexlab{a}}){Zahnle}, {Marley}, \&
  {Fortney}}]{ZahnleEtal2009Soot}
{Zahnle}, K., {Marley}, M.~S., \& {Fortney}, J.~J. 2009{\natexlab{a}}, \apjl,
  submitted, arXiv:0911.0728

\bibitem[{{Zahnle} {et~al.}(2009{\natexlab{b}}){Zahnle}, {Marley}, {Freedman},
  {Lodders}, \& {Fortney}}]{ZahnleEtal2009apjphotochemistry}
{Zahnle}, K., {Marley}, M.~S., {Freedman}, R.~S., {Lodders}, K., \& {Fortney},
  J.~J. 2009{\natexlab{b}}, \apjl, 701, L20, arXiv:0903.1663

\end{thebibliography}

\end{document}